\documentclass[fp,twocolumn]{jpsj3}
\usepackage{txfonts}
\usepackage{color}
\usepackage{amsmath}

\title{Structural Transition in the Hidden Ordered Phase of CeCoSi}

\author{Takeshi Matsumura$^{1}$, Suguru Kishida$^{1}$, Mitsuru Tsukagoshi$^{1}$, Yukihiro Kawamura$^{2}$, Hironori Nakao$^{3}$, and Hiroshi Tanida$^{4}$ }

\inst{$^{1}$Department of Quantum Matter, ADSE, Hiroshima University, Higashi-Hiroshima, Hiroshima 739-8530, Japan \\
$^{2}$Muroran Institute of Technology, Muroran, Hokkaido 050-8585, Japan \\
$^{3}$Photon Factory, Institute of Materials Structure Science, High Energy Accelerator Research Organization, Tsukuba, Ibaraki, 305-0801, Japan \\
$^{4}$Liberal Arts and Sciences, Toyama Prefectural University, Imizu, Toyama 939-0398, Japan\\
}

\abst{
We have performed X-ray diffraction experiments on a single crystalline CeCoSi to investigate the unresolved ordered phase below $T_0 \sim 12$ K. 
We have discovered that a triclinic lattice distortion takes place below $T_0$, which is further modified in the subsequent antiferromagnetic ordered phase. 
The structural domains can be selected by applying a magnetic field, indicating that some electronic ordering exists behind and affects the magnetic anisotropy in the hidden ordered phase below $T_0$. 
The transition at $T_0$, although the order parameter is still unknown, is associated with the maximum in the $c$-axis lattice parameter. 
In magnetic fields along $[1\, 0\, 0]$, the structural transition temperature, named as $T_{\text{s1}}$, deviates from $T_0$ and decreases with increasing the field, whereas $T_0$ increases. 
This shows that the hidden ordered phase without triclinic distortion exists between  $T_{\text{s1}}$ and $T_0$. 
The results for $H \parallel [1\, 1\, 0]$ are also reported. 

(\today)
}

\begin{document}
\maketitle

\section{Introduction}
Inter-ionic interactions between localized $f$-electron orbitals through hybridization with the conduction electrons give rise to a rich variety of ordered phases of multipole moments.~\cite{Kuramoto09} 
In most cases, the order parameters are described by even-parity multipole moments, such as magnetic dipole, electric quadrupole, and magnetic octupole moments, which are inherently allowed in centrosymmetric environments. 
When the spatial inversion symmetry is locally broken at the magnetic ion site, hybridization with the on-site $d$ orbital is allowed, leading to an occurrence of local odd-parity multipoles.~\cite{Hayami18}
Even when a crystal has an inversion center, ferroic order of a cluster-type odd-parity multipole can sometimes be constructed from a staggered ordering of even-parity multipoles, which has been attracting interest as a key concept to understand distinctive magnetoelectric phenomena and unusual ordered phases.~\cite{Suzuki19}

\begin{figure}
\begin{center}
\includegraphics[width=5cm]{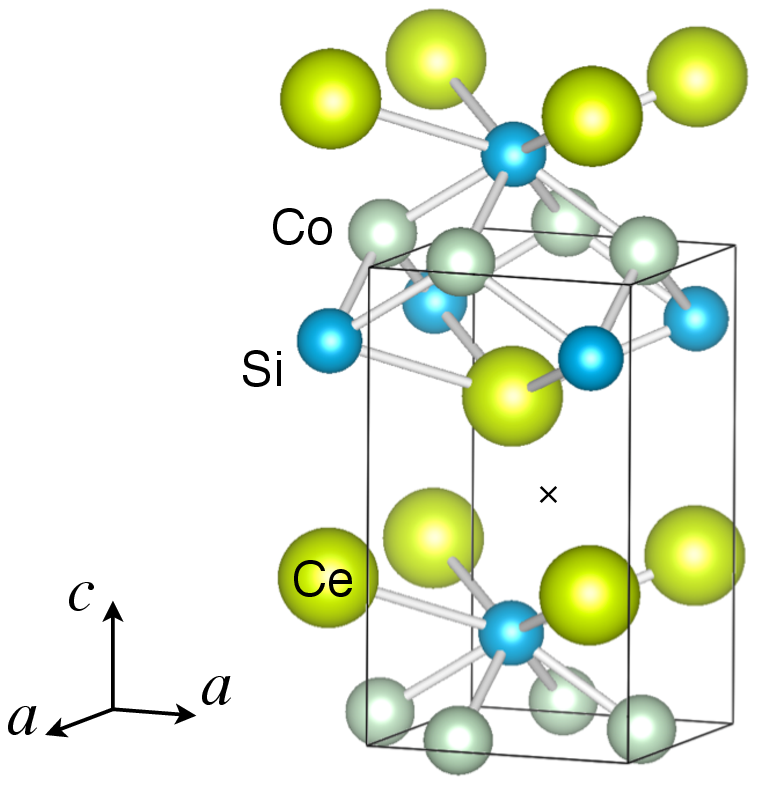}
\caption{(Color online) Crystal structure of CeCoSi.  $a=4.057$ \AA\ and $c=6.987$ \AA\ at room temperature~\cite{Tanida19}. 
A unit cell, represented by the rectangular parallelepiped, contains two formula units. 
The cross at the midpoint of the two Ce atoms in the unit cell represents the inversion center of the crystal. 
The Ce atoms are at the $2c$ site $(1/4, 1/4, z)$ $(z=0.678)$ with the site symmetry $4mm$. 
}
\label{fig:structure}
\end{center}
\end{figure}

CeCoSi, with a tetragonal CeFeSi-type structure (space group $P4/nmm$, No. 129), undergoes an antiferromagnetic (AFM) order at $T_{\text{N}}=9.4$ K~\cite{Chevalier06,Lengyel13,Tanida18,Tanida19,Tanida20}. 
The crystal structure, drawn by using the program VESTA,~\cite{Momma11}  is shown in Fig.~\ref{fig:structure}. 
The magnetic order is described by a propagation vector $\mib{q}=(0, 0, 0)$, where the two Ce atoms in the unit cell, located at noncentrosymmetric sites, are antiferromagnetically coupled.~\cite{Nikitin20}. 
This ordering breaks the global inversion symmetry of the crystal, and gives rise to a cluster-type odd-parity multipole.
The crystalline electric field (CEF) level scheme has been determined by inelastic neutron scattering; the ground state is the $\Gamma_7^{(1)}$ doublet and the $\Gamma_7^{(2)}$ and the $\Gamma_6$ excited doublets are located at 10.49 meV and at 14.1 meV, respectively.~\cite{Nikitin20} 
The ground-state wave function is close to that of the $\Gamma_7$ doublet in cubic CEF, resulting in an almost isotropic character. 
The proposed wave functions well explain the nearly isotropic magnetic properties of CeCoSi.  
It is remarked, however, the size of the ordered moment is less than the value expected from the ground state and is reduced to 0.37 $\mu_{\text{B}}$.~\cite{Nikitin20}

Another noteworthy feature in CeCoSi is the unresolved transition at $T_0 \sim 12$ K. 
Although it has been only detectable as a small anomaly in magnetic susceptibility and specific heat, it is now established to exist as an intrinsic transition by microscopic NQR and NMR measurements.~\cite{Manago21} 
In addition, the transition at $T_0$ is enhanced by applying a high pressure and is connected to the clear anomaly above $\sim 1.5$ GPa with a gap formation in the conduction band.  
Therefore, the ordering at $T_0$ is named as a pressure induced ordered phase (PIOP). 
The transition temperature reaches to an extremely high value of $\sim 36$ K at 1.5 GPa, suggesting an unusual $c$-$f$ hybridization effect~\cite{Lengyel13,Tanida18,Tanida19}. 
The NQR and NMR study concludes that the transition is nonmagnetic in nature.~\cite{Manago21}

It is also important that $T_0$ increases by applying a magnetic field, suggesting a mixing-type quadrupole order by incorporating the excited state. 
It is for the moment a possible candidate for the ordered phase in question as discussed in CeTe, or in an extreme case for YbSb.~\cite{Kawarasaki11,Takaguchi15,Hayashi16,Yamamoto04,Oyamada04}  
However, in a straightforward consideration, such a scenario is difficult to accept because the energy level of the excited state in CeCoSi is too high to have any effect on the phase transition. 
This is the mystery of the hidden order at $T_0$. 
Although it has been discussed in association with the cluster-type odd-parity multipole,~\cite{Yatsushiro20a,Yatsushiro20b} 
it is still unclear whether the lack of local inversion symmetry at the Ce site, or the formation of cluster-type odd-parity multipole, play any role in this ordering, which is one of the main issues in CeCoSi. 

To study the mysterious ordered phase of CeCoSi between $T_{\text{N}}$ and $T_0$ from structural viewpoints, we have performed X-ray diffraction experiments. 
The crystal structure has been considered to maintain the tetragonal symmetry in the whole temperature range as concluded from the results of neutron powder diffraction and Co-NQR~\cite{Nikitin20,Manago21}. 
However, probably due to the higher experimental resolution of single-crystal X-ray diffraction, we have discovered a lattice distortion to take place below $T_0$, which is concluded to be a transition to a triclinic structure. The triclinic distortion is further modified in the AFM ordered phase.   
In addition, the structural domains can be selected by applying a magnetic field, suggesting that an electronic order exists behind and affects the magnetic anisotropy. 
These results pose a new question on the hidden ordered state of CeCoSi.

\section{Experiment}
Single crystals of CeCoSi were grown by Ce/Co eutectic flux method as described in Ref.~\citen{Tanida19}. 
X-ray diffraction experiments were performed at BL-3A of the Photon Factory, KEK, Japan. 
A plate-shaped single-crystalline sample with an as-grown $c$-plane surface, $1.5\times 2.0$ mm$^2$ in area and 0.3 mm in thickness, was mounted in a vertical field 8 Tesla cryomagnet by using Apiezon-N grease so that the $c$-axis was parallel to the horizontal scattering plane.
The $(H, 0, L)$ and $(H, H, L)$ scattering planes were selected by rotating the sample about the $c$-axis, where the field direction was along $[0\, 1\, 0]$ and $[\bar{1}\, 1\, 0]$, respectively.  
We used a two-dimensional Timepix area detector with $256 \times 256$ pixels at a pitch of 55 $\mu$m, which covered approximately two degrees each for the horizontal ($2\theta$) and vertical ($\chi$) directions. 
The total intensity of the Bragg peak was measured by an oscillation method, where the crystal angle ($\theta$) was rotated by approximately two degrees about the Bragg peak. 
We used an X-ray beam with a wavelength of 1.385 \AA \, (8.952 keV) in a nonresonant region. 
We also used a laboratory-based rotating anode X-ray source (Cu-K$\alpha$) and a two-axis diffractometer to measure the temperature ($T$) dependence of the lattice parameters between 3 K and 300 K for the same single crystal. 

\section{Results and Analysis}
\subsection{Temperature and magnetic-field dependences}
\begin{figure}
\begin{center}
\includegraphics[width=8.5cm]{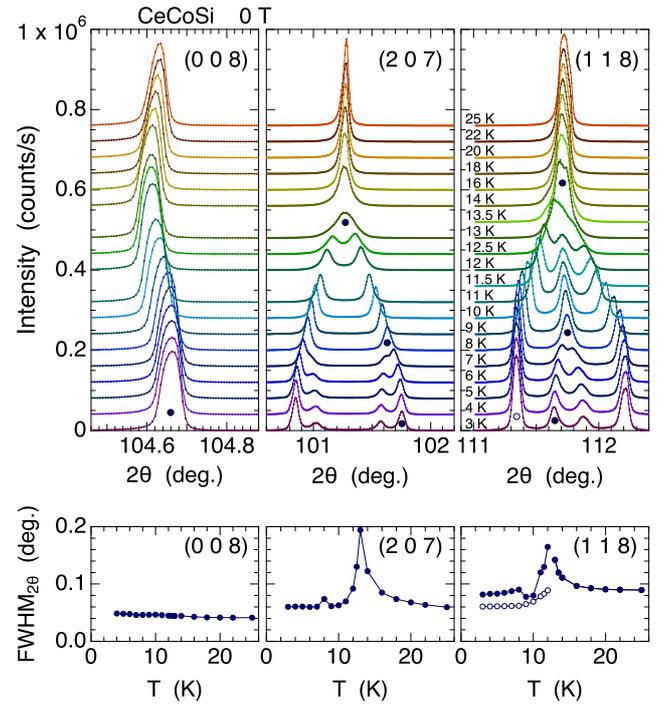}
\caption{(Color online) Top: Scattering angle dependences of the intensity of the (0 0 8), (2 0 7), and (1 1 8) Bragg reflections at zero field between 3 K and 25 K. Note the different scale of the horizontal axis for (0 0 8). 
Bottom: Temperature dependences of the full width at half maximum (FWHM) of the Bragg peak indicated by the mark (closed circle) in the top panels. 
For (1 1 8), the FWHM for another peak indicated by the open circle is also shown. 
The marks on the (2 0 7) and (1 1 8) peaks at 8 K and 13 K, below which the peak is split, represent that the peak is treated as a single peak in evaluating the FWHM at and above these temperatures.
 }
\label{fig:tthscans0T}
\end{center}
\end{figure}

Figure~\ref{fig:tthscans0T} shows the scattering angle ($2\theta$) dependences of the intensity of the (0 0 8), (2 0 7), and (1 1 8) fundamental Bragg reflections at temperatures ranging from 3 K to 25 K. The original two dimensional data have been converted to the one dimensional $2\theta$ dependence by integrating the pixel intensity along the vertical direction~\cite{SM}. 
As it is clearly observed, the (0 0 8) peak does not show any splitting and the peak position exhibits a smooth $T$-dependence with a minimum at 13 K. 
The (2 0 7) and (1 1 8) peaks, on the other hand, exhibit clear splittings below 13 K. 
The (2 0 7) peak splits into two peaks below 13 K, both of which further split into two peaks below 8 K, which is slightly lower than $T_{\text{N}}$.  
The (1 1 8) peak splits into three peaks below 13 K and the central peak splits into two peaks below 8 K. 

The $T$-dependences of the full width at half maximum (FWHM) of the Bragg peaks, obtained from the fits to multiple squared Lorentzian functions, are also shown in the bottom panels of Fig.~\ref{fig:tthscans0T}. These are the FWHMs of the representative peak indicated by the mark in the top panels. 
For (0 0 8), the peak profiles at all temperatures were treated as a single peak. 
For (2 0 7) and (1 1 8), the profiles for $T \ge 13$ K, as indicated by the marks (closed circle on the data at 13 K in the top panels), were treated as a single peak. 
The profiles for $8 \le T \le 12.5$ K were treated as double peaks for (2 0 7) and triple peaks for (1 1 8). 
For  $T \le 7$ K the profiles were treated as quadruple peaks. 
The broad profile at 12.5 K for (1 1 8) was excluded from the FWHM analysis because it was difficult to resolve the three peaks. 
The FWHM of the (0 0 8) peak does not exhibit any noticeable anomaly since no splitting takes place. The FWHM of the (2 0 7) and (1 1 8) peaks, by contrast, exhibits a significant increase with decreasing $T$ toward 13 K, implying an increase in the structural instability. 
Below the transition at 13 K, the width decreases with decreasing $T$ and again exhibits a weak increase at 8 K. 


\begin{figure}
\begin{center}
\includegraphics[width=8cm]{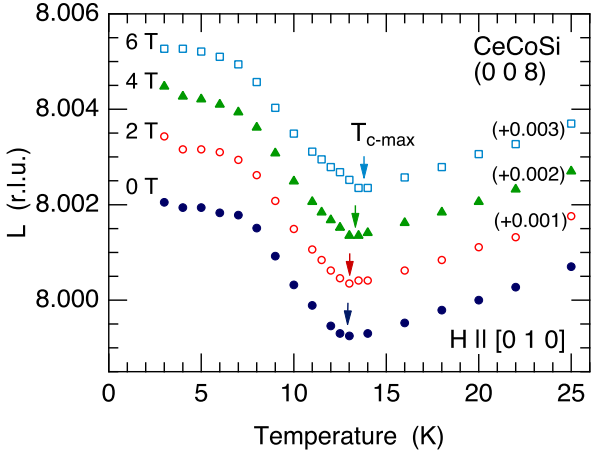}
\caption{(Color online)
Temperature dependence of the peak position of the (0 0 8) reflection, $L$ in $(H, K, L)$, expressed in the reciprocal lattice unit of the tetragonal lattice. 
The temperature indicated by the arrow is named as $T_{c\text{-max}}$, where the $c$-axis lattice parameter takes the maximum. 
Magnetic fields are applied along the $[0\, 1\, 0]$ direction. The data for 2 T, 4 T, and 6 T are shifted by 0.001, 0.002, and 0.003, respectively. 
The reciprocal lattice is defined so that $L=8$ at 20 K. 
}
\label{fig:Tdep008L}
\end{center}
\end{figure}

Figure~\ref{fig:Tdep008L} shows the $T$-dependence of the peak position, $L$ in ($H, K, L$), of the (0 0 8) reflection in the reciprocal lattice of the original tetragonal unit cell. Since the minimum in $L$ corresponds to the maximum in the lattice parameter $c$, the temperature at which $L$ takes the minimum is named as $T_{c\text{-max}}$. 
When a magnetic field is applied along $[0\, 1\, 0]$, $T_{c\text{-max}}$ shifts to higher temperatures. 
It is noted that the specific heat of the sample used in this study shows the anomaly of $T_0$ at 13 K~\cite{SM}. 
These results show that $T_{c\text{-max}}$ corresponds to $T_0$.  

It is also anomalous that the lattice parameter $c$ decreases with increasing temperature above $T_0$, indicating a negative thermal expansion.  
This unusual behavior persists up to the room temperature~\cite{SM}. 
The lattice parameter $a$, on the other hand, exhibits a normal $T$-dependence as it increases almost linearly with $T$. 
It is noted that the relation $\Delta c/c\approx -0.75\Delta a/a$ holds above $T_{c\text{-max}}$. 
In view of the two-dimensional nature of the crystal structure, where the rigid Co-Si layers seem to be weakly coupled along the $c$-axis~\cite{Tanida19}, the negative thermal expansion along the $c$-axis suggests that the total thermal expansion is mostly determined by the in-plane bonding; i.e., when the $ab$-plane contracts, the lattice consequently expands along the $c$-axis. 
This may be associated with the distinctive pressure dependence of $c/a$ in CeCoSi~\cite{Kawamura20,Kawamura22}.

\begin{figure}
\begin{center}
\includegraphics[width=8cm]{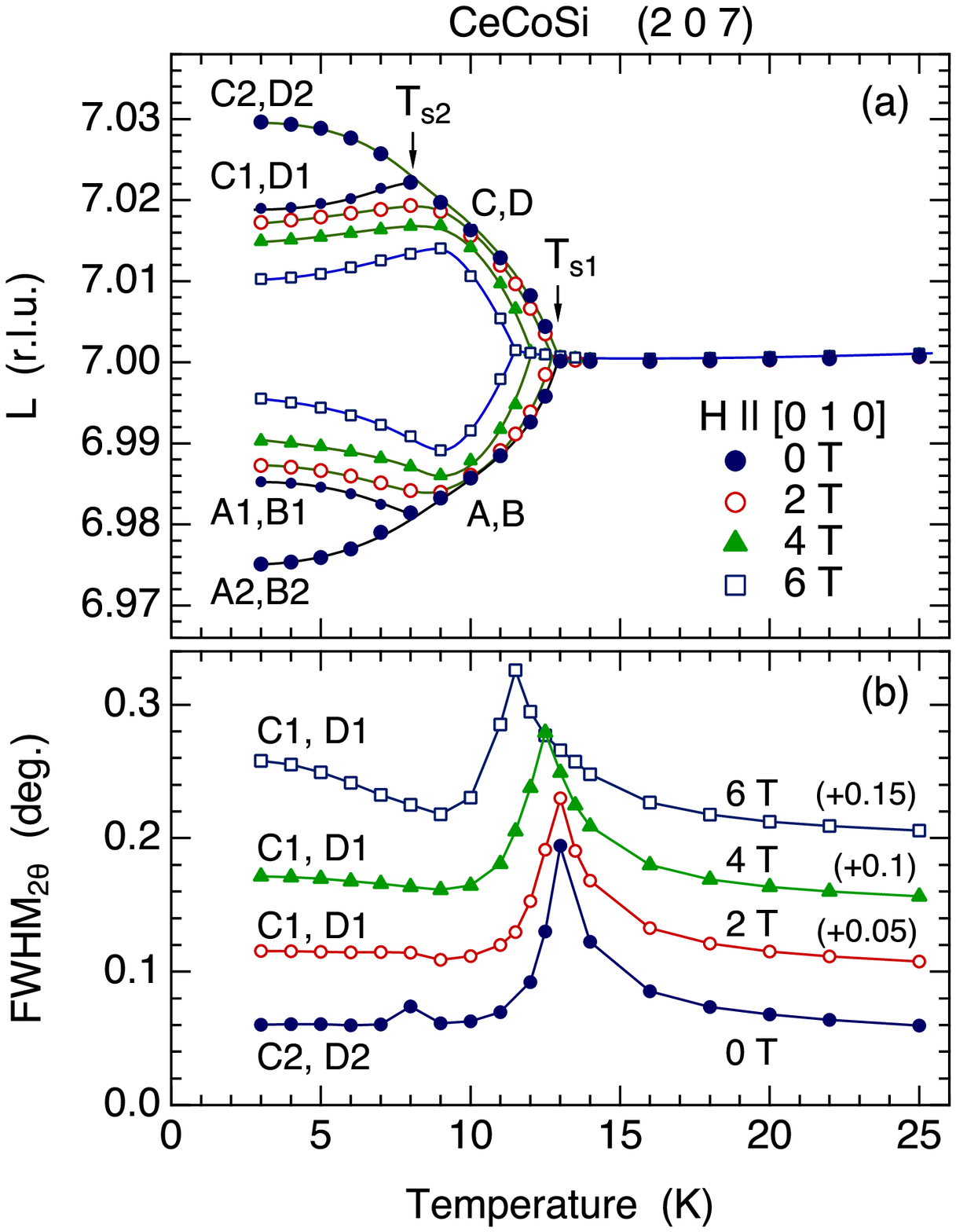}
\caption{(Color online)
(a) Temperature dependences of the peak position of the (2 0 7) reflection, $L$ in $(H, K, L)$, expressed in the reciprocal lattice unit of the tetragonal lattice. Magnetic fields are applied along the $[0\, 1\, 0]$ direction. The labels represent the domain assignments shown in Fig.~\ref{fig:triclinic}. 
(b) Temperature dependences of the FWHM for the peak corresponding to the structural domain represented by the label. 
The data for 2 T, 4 T, and 6 T are shifted by the value shown in the parentheses
}
\label{fig:Tdep207L}
\end{center}
\end{figure}

Figure~\ref{fig:Tdep207L}(a) shows the $T$-dependences of the peak position, $L$ in ($H, K, L$), of the (2 0 7) reflection in magnetic fields applied along the $[0\, 1\, 0]$ direction.~\cite{SM} 
As demonstrated in Fig.~\ref{fig:tthscans0T}, the (2 0 7) peak exhibits a clear splitting below $T_0=13$ K at zero field into two peaks, indicating an occurrence of a structural transition at $T_0$. 
In magnetic fields, the structural transition temperature, which we name as  $T_{\text{s1}}$, decreases with increasing the field.  
On the other hand, $T_{c\text{-max}}$ $(=T_0)$ increases with increasing the field, which is also observed as a minimum in $L$ in (2 0 7). 
Below $T_{c\text{-max}}$ the peak position $L$ increases with decreasing $T$ down to $T_{\text{s1}}$, where the peak splits into two. 
This result shows that $T_{\text{s1}}$ separates from $T_0$ in magnetic fields, whereas they coincide at zero field. 
As shown in Fig.~\ref{fig:Tdep207L}(b) for a representative peak, the FWHM increases with decreasing $T$ toward $T_{\text{s1}}$, implying an increase in the structural instablility at all magnetic fields (see supplemental material for the original peak profile).~\cite{SM} 
After exhibiting a sharp peak at $T_{\text{s1}}$, the FWHM decreases toward the resolution limit of 0.05$^{\circ}$ for (2 0 7) with decreasing $T$.   

To deduce the FWHM, the peak profile at $T_{\text{s1}}$ (e.g. 13 K for 0 T) is treated as a single peak and that just below $T_{\text{s1}}$ (e.g. at 12.5 K for 0 T) is treated as double peaks. 
It is noted that, at 0 T for example, the FWHM $\sim 0.2^{\circ}$ at 13 K is almost equal to the peak split at 12.5 K. 
It is therefore not possible to distinguish experimentally whether the peak profile at 13 K is double peaked with a split by $\sim 0.1^{\circ}$ or is single peaked with a FWHM of $\sim 0.2^{\circ}$ (both can fit the data). 
The purpose of plotting the FWHM in this paper, whichever the case may be, is to show and emphasize the tendency of the peak splitting, i.e. the existence of lattice instability at the structural transition. 

In the AFM ordered phase, the double peak splits into four peaks at zero field, indicating an occurrence of a second structural transition. 
Since the transition temperature of 8 K appears to be slightly lower than $T_{\text{N}}$, we named the temperature as $T_{\text{s2}}$ to distinguish it from $T_{\text{N}}$. 
At $T_{\text{s2}}=8$ K, the FWHM of the Bragg peak exhibits a weak increase at 0 T. 
As explained above, this suggests a structural instability taking place below $T_{\text{N}}=9.4$ K before it is fixed below $T_{\text{s2}}=8$ K and the peak starts to split. 
It is also noted that the $T$-dependence of the peak split at 0 T shown in Fig.~\ref{fig:Tdep207L}(a) seems to close at 8 K by extrapolation and not at 9 K. 
This point will be discussed later in Sec.~\ref{sec:PD}. 

By applying a field of 2 T, the peak splitting below $T_{\text{s2}}$ soon disappears because the structural domains corresponding to the inner peaks are selected. 
The domain selection below $T_{\text{N}}$ above 2 T, or in other words a structural transition across $T_{\text{N}}$, is observed as the maximum in the peak split at around 9 K, which seems to coincide with $T_{\text{N}}$. 
The FWHM does not exhibit a cusp anomaly in magnetic fields above 2 T after the preferable domains are selected. 
With respect to the data at 6 T, the FWHM increases with decreasing $T$ down to the lowest temperature of 3 K. 
This reflects the approach to another boundary at around 8 T as explained next. 

\begin{figure}
\begin{center}
\includegraphics[width=8cm]{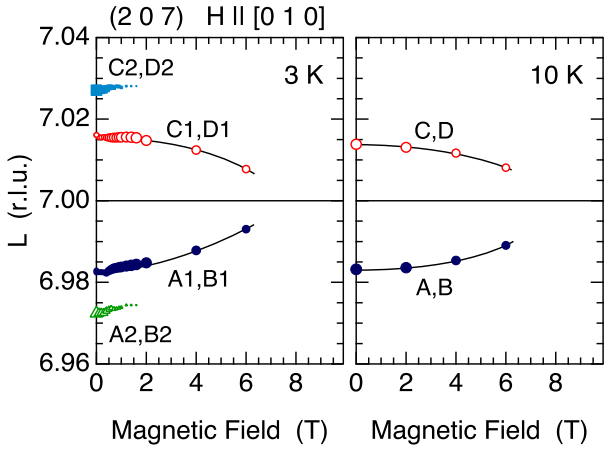}
\caption{(Color online)
Magnetic field dependence of the peak position of the (2 0 7) reflection, $L$ in $(H, K, L)$, at 3 K and 10 K, expressed in the reciprocal lattice unit of the tetragonal lattice. The size of the marks represent the peak intensity. 
 }
\label{fig:Hdep207L}
\end{center}
\end{figure}

Magnetic field dependence of the peak position $L$ of the (2 0 7) reflection for $H \parallel [0\, 1\, 0]$ is shown in Fig.~\ref{fig:Hdep207L}.~\cite{SM}
At 3 K, the outer peaks soon disappear by applying a field of $\sim 1$ T and the intensity is transferred to the inner peaks, indicating that the preferable structural domains are selected by the magnetic field. 
The intensity of the outer peaks almost disappear at 2 T. 
The width of the split between the inner peaks gradually decrease above 2 T. 
The split peaks are expected to merge back into the original central peak at around 8 T. 
This is also the case at 10 K. The critical field at 3 K seem to be lower than that at 10 K. 

\begin{figure}
\begin{center}
\includegraphics[width=8cm]{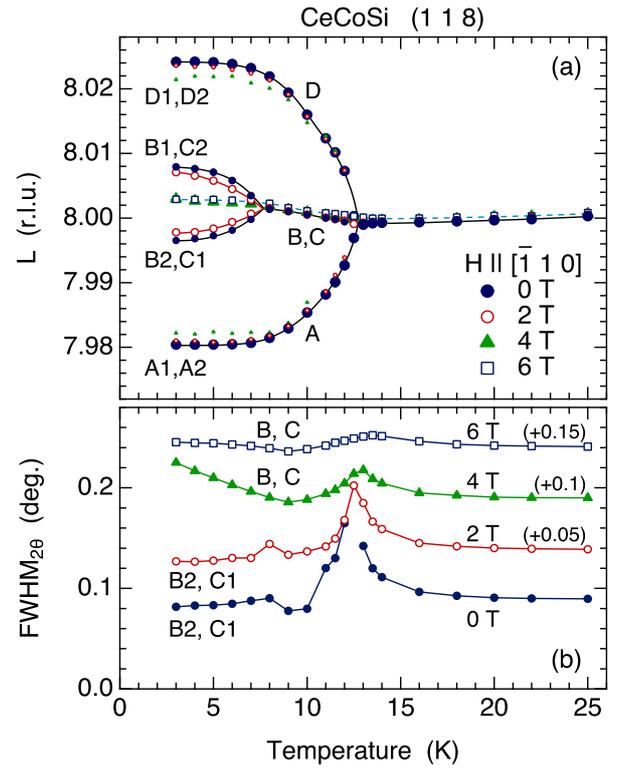}
\caption{(Color online) 
(a) Temperature dependences of the peak position of the (1 1 8) reflection, $L$ in $(H, K, L)$, expressed in the reciprocal lattice unit of the tetragonal lattice. Magnetic fields are applied along the $[\bar{1}\, 1\, 0]$ direction. The labels represent the domain assignments shown in Fig.~\ref{fig:triclinic}. 
(b) Temperature dependences of the FWHM for the peak corresponding to the structural domain represented by the label. 
The data for 2 T, 4 T, and 6 T are shifted by the value shown in the parentheses. 
 }
\label{fig:Tdep118L}
\end{center}
\end{figure}

Next, we rotated the crystal by $45^{\circ}$, changed the scattering plane to $(H, H, L)$, and investigated the (1 1 8) reflection. 
Figure~\ref{fig:Tdep118L}(a) shows the $T$-dependence of the peak position, $L$ in ($H, K, L$), of the (1 1 8) reflection in magnetic fields applied along the $[\bar{1}\, 1\, 0]$ direction. 
First, at zero field, the (1 1 8) peak splits into three below $T_{\text{s1}}$ and the central peak splits into two below $T_{\text{s2}}$. 
In magnetic fields, the split outer peaks become weak because of the domain selection.~\cite{SM}  
At 6 T, it was difficult to determine the ($H, K, L$)-index because the peak intensity was too weak. 
From the $T$-dependence of the peak split at 2 T, $T_{\text{s1}}$ does not seem to change by the field, which is different from the result of (2 0 7) for $H\parallel [0\, 1\, 0]$. 
On the other hand, the peak position $L$ of the central peak exhibits a minimum at $T_0$ and increases with decreasing $T$ below $T_0$. 
It seems difficult to distinguish whether $T_{\text{s1}}$ deviates from $T_0$ or not from the peak split because the domain population is mostly transferred to the central peak and the intensities of the split peaks become very weak above 4 T. 

Figure~\ref{fig:Tdep118L}(b) shows the $T$-dependences of the FWHM of the central peaks that remain up to high fields.~\cite{SM} 
Although it was difficult to follow the field dependence of $T_{\text{s1}}$ from the peak split as described above, the FWHM exhibits a cusp anomaly even at 6 T where the split peaks are difficult to observe. 
The increase in FWHM, reflecting the structural instability, strongly suggests that the transition at $T_{\text{s1}}$ still exists even at 6 T and that $T_{\text{s1}}$ increases with increasing the field, i.e., $T_{\text{s1}}=T_0$ for $H\parallel [\bar{1}\, 1\, 0]$. 
Since the increase in FWHM reflects the tendency of the peak splitting, the anomaly becomes weaker as the intensities of the split peaks transfer to the central peak due to the domain selection by the applied field. This is a different situation for $H\parallel [\bar{1}\, 1\, 0]$ than for $H\parallel [0\, 1\, 0]$ in Fig.~\ref{fig:Tdep207L}(b). 
At $T_{\text{s2}}=8$ K the FWHM again exhibits a weak cusp anomaly at 0 T and 2 T, reflecting the lattice instability taking place in the AFM phase. 
Above the phase boundary at 4 T, which is explained next, the cusp anomaly disappears. 
At the boundary of 4 T, the FWHM increases with decreasing $T$ below 9 K, reflecting the increase in structural instability by approaching the phase boundary. 

\begin{figure}
\begin{center}
\includegraphics[width=8cm]{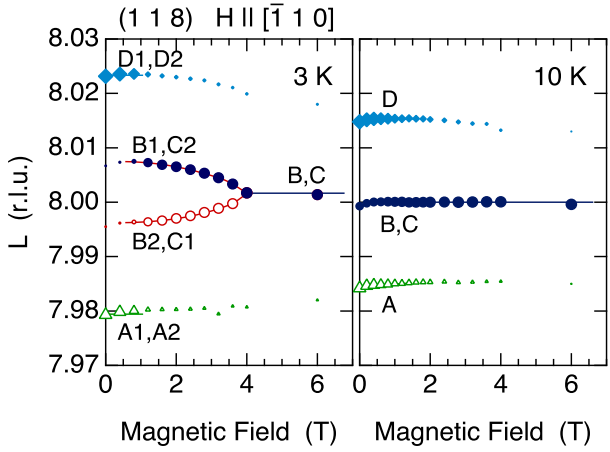}
\caption{(Color online)
Magnetic field dependence of the peak position of the (1 1 8) reflection, $L$ in $(H, K, L)$, at 3 K and 10 K, expressed in the reciprocal lattice unit of the tetragonal lattice. The size of the marks represent the peak intensity. 
 }
\label{fig:Hdep118L}
\end{center}
\end{figure}

Magnetic field dependence of the peak position $L$ of the (1 1 8) reflection for $H \parallel [\bar{1}\, 1\, 0]$ is shown in Fig.~\ref{fig:Hdep118L}.~\cite{SM}
At 3 K, as for (2 0 7), the intensity of the outer peaks is rapidly transferred to the inner peaks at around 1 T, indicating the domain selection by the field. 
The width of the peak split between the inner peaks gradually decreases and the split peaks merge back into the central peak at 4 T. 
The intensity of the outer peaks weakly remains at 4 T, which further decreases at 6 T to a negligible level, but it still appears to remain finite. 
Also at 10 K, the intensity of the outer peak shows a steep decrease at around 1 T and a gradual decrease up to 4 T. The intensity still appears to remain finite even at 6 T. 
The decreased intensity is transferred to the central peak. 

\subsection{Triclinic distortion}
\begin{figure}
\begin{center}
\includegraphics[width=8cm]{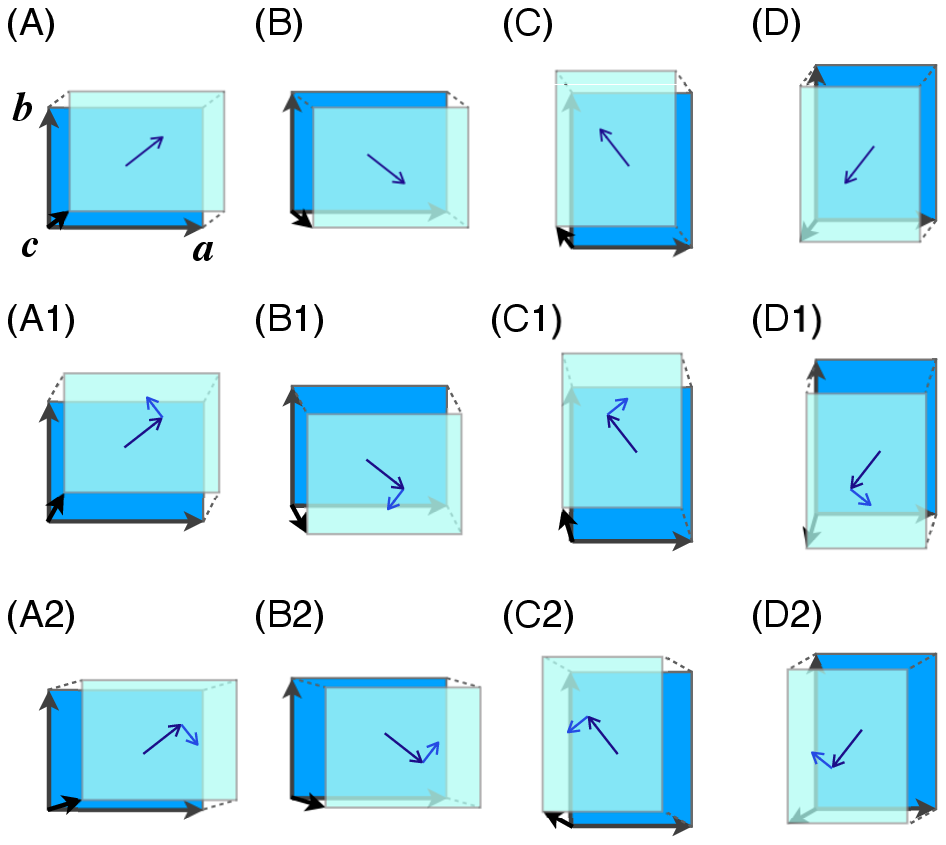}
\caption{(Color online)
A model of the triclinic structure after the structural transition. 
A, B, C, and D represent the four types of structural domains below $T_{\text{s1}}$.  
The arrow on the upper $ab$-plane represents the direction of the shift of the plane from the tetragonal lattice. 
A1, A2 \textit{etc.} are the domains below $T_{\text{s2}}$. 
The short arrow on the upper $ab$-plane represents the direction of the secondary shift of the plane below $T_{\text{s2}}$. 
 }
\label{fig:triclinic}
\end{center}
\end{figure}

Let us analyze the crystal structure at zero field to consistently explain the peak splitting of the (0 0 8), (2 0 7), and (1 1 8) reflections shown in Figs.~\ref{fig:Tdep008L}, \ref{fig:Tdep207L}, and \ref{fig:Tdep118L}, respectively. 
No splitting in the (0 0 8) reflection shows that the direction of the reciprocal lattice vector $c^*$ does not change. 
The splitting of the (2 0 7) and (1 1 8) Bragg peaks into two and three along $L$, respectively, can be explained by considering that the $c$-axis tilts to the $[1\,1\,0]$ and the equivalent directions, consequently forming a monoclinic lattice. 
Finally, to explain the splitting of $H$, it is necessary to introduce different lattice parameters for $a$ and $b$, indicating that the lattice is distorted to triclinic.~\cite{SM} 
The distorted lattice in the temperature range $T_{\text{s2}} < T < T_{\text{s1}}$ is illustrated in Fig.~\ref{fig:triclinic}(A) and its equivalent domains are shown in Figs.~\ref{fig:triclinic}(B), (C), and (D). 
With respect to the domain-A at 10 K, the relative change in the lattice parameters with respect to those at 20 K are estimated to be 
$\Delta a/a=5.0 \times 10^{-4}$, $\Delta b/b= -5.0 \times 10^{-4}$, and $\Delta c/c=-2.2 \times 10^{-4}$. 
The triclinic angles are $\alpha= 89.64^{\circ}$, $\beta= 89.64^{\circ}$, and $\gamma= 90.0^{\circ}$. 
The volume contraction is calculated to be $\Delta V/V =-2.6 \times 10^{-4}$, which is consistent with fact that $T_0$ increases by applying an external pressure.~\cite{Lengyel13,Tanida18,Tanida19,Manago21}
Although the angle $\gamma$ is not necessarily equal to $90^{\circ}$ in the triclinic lattice, the deviation is too small (less than $0.05^{\circ}$) to determine within the accuracy of the present measurement. 
All the data were consistently explained by assuming $\gamma= 90.0^{\circ}$.

The peak split below  $T_{\text{s2}}$ can be explained by considering an additional tilt of the $c$-axis to the directions perpendicular to the first one below $T_{\text{s1}}$. For each domain from A to D, there arise two domains, which are named A1, A2, etc., resulting in eight domains in total. 
The peak assignments to these domains are given in Figs.~\ref{fig:Tdep207L}, \ref{fig:Hdep207L}, \ref{fig:Tdep118L}, and \ref{fig:Hdep118L}. 

In magnetic fields of $H \parallel [0\, 1\, 0]$, the A1, B1, C1, and D1 domains are selected and the A2, B2, C2, and D2 domains disappear.~\cite{SM}  
At 3 K in the AFM phase, the intensities of the outer peaks (A2, B2, C2, D2) rapidly decrease by applying a field of $\sim 1$ T and almost disappear at 2 T.  The intensities transfer to the inner peaks (A1, B1, C1, D1). 
With further increasing the field, the width of the split between the inner peaks gradually decreases, and the peaks are expected to merge back into the original central peak at around 8 T.

In magnetic fields of $H \parallel [\bar{1}\, 1\, 0]$, the domains B and C are selected and the domains A and D disappear approximately at 1 T.  
At 3 K in the AFM phase, the width of the peak split between the inner peaks of the domains B and C gradually decrease and merge back into the central peak at 4 T. 
It is noted, however, as shown by the tiny marks in Figs.~\ref{fig:Tdep118L} and \ref{fig:Hdep118L}, the intensities of the outer peaks of the domains A and D weakly remain up to 4 T. Although they become very weak at higher fields, they still survive even at 6 T (see supplemental material). 
From these results of domain selection, we can see that the structural domains are preferred in which the tilt of the $c$-axis has more parallel component to the applied field. This is a common nature irrespective of the temperature above or below $T_{\text{N}}$.

\subsection{Magnetic field vs temperature phase diagram}
\label{sec:PD}
\begin{figure}
\begin{center}
\includegraphics[width=8cm]{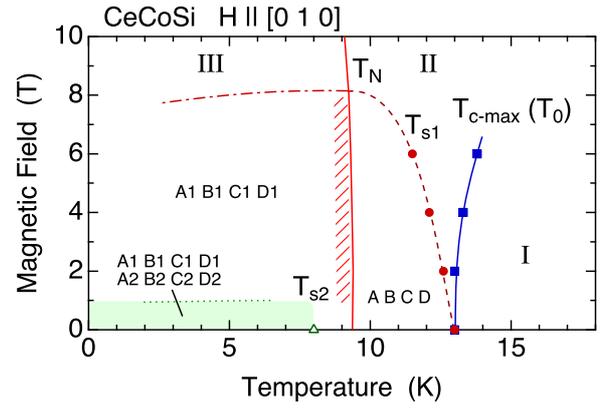}
\caption{(Color online)
Magnetic-field vs temperature phase diagram of CeCoSi for $H \parallel [0\, 1\, 0]$ constructed from the results of (2 0 7) and (0 0 8) reflections. 
$T_{c\text{-max}}$ represents the temperature where the lattice parameter $c$ takes the maximum, which is equal to $T_0$. 
Vertical solid line represents $T_{\text{N}}$ taken from the literature.~\cite{Tanida19} 
$T_{\text{s1}}$ and $T_{\text{s2}}$ are the temperatures below which the (2 0 7) peak splits into two and four peaks, respectively. 
The dashed line connecting $T_{s1}$ is a guide for the eyes. 
The horizontal dotted line at 1 T represents the boundary of domain selection. See Fig.~\ref{fig:triclinic} for the name of the triclinic domains. 
The shaded area around 9 K represents the structural crossover across $T_{\text{N}}$. 
 }
\label{fig:PD010}
\end{center}
\end{figure}

From the above results, we have constructed a phase diagram from the structural viewpoint. 
The result for $H \parallel [0\, 1\, 0]$ is shown in Fig.~\ref{fig:PD010}. 
First, the onset temperature of the hidden order, $T_0$, is equal to $T_{c\text{-max}}$ and increases with increasing the filed. 
This is consistent with the previous reports.~\cite{Tanida19,Tanida20,Manago21} 
Second, a structural transition takes place at $T_{\text{s1}}$, which is equal to $T_0$ at zero field and decreases with increasing the field. 
Below $T_{\text{s1}}$, the (2 0 7) peak splits into two peaks. 

As shown in Fig.~\ref{fig:Tdep207L}(b), the FWHM exhibits a significant increase at $T_{\text{s1}}$. 
This is considered to reflect an enhancement of the triclinic structural instability on approaching $T_{\text{s1}}$. 
At temperatures above $T_{\text{s1}}$, although the spatial and time average of the structure must still be tetragonal, or orthorhombic in magnetic fields in a strict sense, short-ranged or instantaneous fluctuation of the triclinic phase is enhanced on approaching $T_{\text{s1}}$, where the phase transition finally takes place and the Bragg peak splits. 

At zero field, the (2 0 7) peak further splits into four peaks below $T_{\text{s2}}$, where the triclinic structure is modified by the appearance of the AFM order.  
The result that $T_{\text{s2}}$ appears to be lower than $T_{\text{N}}$ by $\sim 1$ K at 0 T can possibly be considered as reflecting the similar kind of structural instability of the low-$T$ triclinic phase, which becomes significant just below $T_{\text{N}}$ on entering the AFM phase. 
For example, when the AFM order takes place in the triclinic domain-A, there can arise a structural fluctuation between the two domains of A1 and A2, which consequently results in a slightly lower transition temperature of $T_{\text{s2}}$ than $T_{\text{N}}$. 
The enhanced instability is reflected in the weak increase in the FWHM at 8 K at 0 T as shown  in Fig.~\ref{fig:Tdep207L}(b). 

Above 1 T, as represented by the horizontal dotted line, the preferable structural domains below $T_{\text{N}}$ (A1, B1, C1, D1) are selected, leaving two peaks to remain. The two peaks are expected to vanish at around 8 T, which is expected to be connected to $T_{\text{s1}}$ above $T_{\text{N}}$. 
When the preferable domains are selected above 1 T, a structural transition (or a crossover) from phase II ($T_0 > T>T_{\text{N}}$) to III ($T<T_{\text{N}}$) is observed at around 9 K, which is in agreement with $T_{\text{N}}$. This may be because the structural fluctuation is suppressed by selecting the preferable structural domain by the applied field. 

\begin{figure}
\begin{center}
\includegraphics[width=8cm]{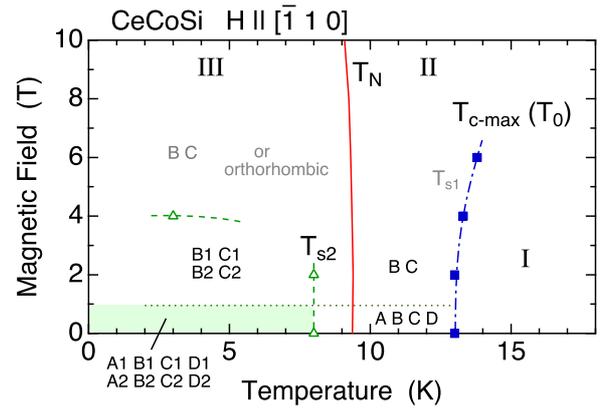}
\caption{(Color online) 
Magnetic-field vs temperature diagram for $H \parallel [\bar{1}\, 1\, 0]$ constructed from the results of (1 1 8) reflection. 
It has not been clarified whether the B and C triclinic domains persist up to high fields above 4 T with $T_{\text{s1}}=T_0$ or the high field phase is orthorhombic without structural transition at $T_{\text{s1}}$. 
}
\label{fig:PD110}
\end{center}
\end{figure}

The phase diagram for $H \parallel [\bar{1}\, 1\, 0]$ constructed from the available data, though not comprehensive yet, is shown in Fig.~\ref{fig:PD110}. 
For $H \parallel [\bar{1}\, 1\, 0]$, $T_0$ is almost equal to that for $H \parallel [0\, 1\, 0]$ as observed in the maximum of the lattice parameter $c$.  
The second structural transition at $T_{\text{s2}}$ still exists even in a magnetic field of 2 T as shown in Fig.~\ref{fig:Tdep118L}(a). 
This may be because the structural fluctuation remains in magnetic fields for $H \parallel [\bar{1}\, 1\, 0]$ even above the initial domain selection at 1 T. 
For example, when the AFM order takes place in the triclinic domain-B at 2 T, there remains a fluctuation between the domains B1 and B2, which causes the second structural transition at $T_{\text{s2}}$ and causes the peak splitting within the preferred domain. 
We consider that this is reflected in the increase in the FWHM at $T_{\text{s2}}$ as shown in Fig.~\ref{fig:Tdep118L}(b) for 0 T and 2 T. 

There is also a boundary at 4 T as observed in the disappearance of the peak split at 3 K in the central peak shown in Fig.~\ref{fig:Hdep118L}. 
This boundary is also reflected in the increase in the FWHM with decreasing $T$ below 9 K at 4 T as shown in Fig.~\ref{fig:Tdep118L}(b).
In magnetic fields above 4 T, the B and C structural domains are selected and the intensities from the A and D domains significantly decrease.  
It is noted that, since the central peak of the (1 1 8) reflection corresponding to the B and C domains does not split, it is not clear whether the B and C triclinic domains persist or they come back to the tetragonal structure, or an orthorhombic structure in a strict sense in magnetic fields.~\cite{SM} 
We cannot distinguish the two possibilities. 
However, judging from the fact that the weak intensities from the A and D domains remain above 4 T, indicating the domain selection and not the change in the lattice type and the symmetry, it is inferred that the B and C domains continue to exist up to higher fields. 
In addition, also from the result that the FWHM of the central peak exhibits a cusp anomaly at $T_0$ at 4 T and 6 T, reflecting the peak split due to the appearance of the A and D domains, the transition from the tetragonal to the triclinic domains of B and C is also likely to exist at high fields above 4 T. 
If this is the case, $T_{\text{s1}}=T_0$ is concluded for $H \parallel [\bar{1}\, 1\, 0]$. 

However, since the main peak of the (1 1 8) reflection at the center above 4 T is unambiguously a single peak, from which we cannot extract any evidence for the triclinic distortion. 
If we neglect the tiny peaks for the A and D domains as residual signals, there arise also a possibility that the structure above 4 T for $H \parallel [\bar{1}\, 1\, 0]$ could be orthorhombic. 
In this case, the weak anomaly in FWHM at $T_{\text{s1}}$ is regarded as extrinsic. It is concluded that $T_{\text{s1}}$ does not exist above 4 T for $H \parallel [\bar{1}\, 1\, 0]$. 
These two possibilities above 4 T are written by gray letters in the phase diagram in Fig.~\ref{fig:PD110}. 
The phase diagram for $H \parallel [\bar{1}\, 1\, 0]$ needs to be studied in more detail, which must be different from that for $H \parallel [0\, 1\, 0]$.

\section{Discussions}
Since the magnetic field itself does not have a direct interaction with the lattice, there should be some magnetic anisotropy originating from the electronic system in the triclinic phase, through which the structural domains are manipulated. 
Such an example can be found in the domain switching in the hexagonal YbAl$_3$C$_3$ and DyAl$_3$C$_3$, where the structural domains in the low temperature orthorhombic phase can be manipulated by the magnetic field even in the paramagnetic state probably through the very weak magnetic anisotropy.~\cite{Matsumura15} 
Also in the present case of CeCoSi, the magnetic anisotropy in the $ab$-plane is considered to be very small since the CEF ground state is nearly isotropic.~\cite{Nikitin20}  
Although it is speculated that the ferroquadrupole order of $O_{yz}+O_{zx}$ corresponding to the triclinic distortion gives rise to such a magnetic anisotropy, it still requires further study if such an ordering could be possible by considering the inter-orbital hybridization due to the noncentrosymmetry. 

If the lattice is uniformly distorted in the triclinic structure, probably with the space group $P\bar{1}$ where all the atoms are at the general $2i$ site, there should arise no splitting in the NQR spectrum, but with a shift below $T_0$, which is consistent with the experimental observation.  
To explain the peak splitting in NMR in magnetic fields, on the other hand, it is necessary to take into account an antiferro-type induced moment at the Co site.~\cite{Manago21}  
Even if we simply consider a ferroquadrupole order for the structural transition, it is still necessary to consider some antiferro-type order taking place below $T_0$.~\cite{Yatsushiro20b}

Therefore, the phase transition at $T_0$ is still in question. 
It should be associated with the Ce-$4f$ and a Ce-Ce interaction, possibly of nonmagnetic origin involving some orbital nature, since the order is fragile to La substitution.~\cite{Tanida18} 
In particular, the relation with the triclinic distortion is mysterious. 
For $H \parallel [0\, 1\, 0]$, $T_0$ and $T_{\text{s1}}$ separate with increasing the field. 
The temperature region $T_{\text{s1}} < T < T_0$ is a purely hidden ordered phase without triclinic distortion, but with a lattice contraction along the $c$-axis and with some antiferro-type order of orbital nature. 
We note that the phase boundary at $T_{\text{s1}}$ in magnetic fields has also been revealed recently by the bulk property measurements.~\cite{Hidaka22}  

One of the possibilities of the crystal structure of the purely hidden ordered phase for $H \parallel [0\, 1\, 0]$, which is induced by the underlying electronic order, could be a monoclinic one with the space group $P112/n$ (No. 13), a subgroup of an orthorhombic space group $Pmmn$ (No. 59), where $Pmmn$ is a subgroup of the tetragonal $P4/nmm$ (No. 129).~\cite{SM} 
Although there has been no experimental evidence for the symmetry lowering from tetragonal to orthorhombic in the paramagnetic phase above $T_0$, it may be assumed that the symmetry of the crystal is reduced to orthorhombic in a magnetic field applied along the $[0\,1\,0]$ direction. 
A possible consequence of the antiferro-type hidden order at $T_0$ could be an $O_{xy}$-type ferroquadrupole order, resulting in the symmetry lowering from orthorhombic to monoclinic. 
Since no peak splitting is observed in the region $T_{\text{s1}} < T < T_0$, the monoclinic distortion is expected to be negligibly small. 
In this phase, the $c$-axis is still perpendicular to the $ab$-plane. The $O_{yz}$ and $O_{zx}$ moments are induced below $T_{\text{s1}}$ with the reduction to $P\bar{1}$. 
In our observation, the transition at $T_0$ is only reflected in the maximum of the $c$-axis lattice parameter. 
At zero field, this electronic order and the triclinic distortion to $P\bar{1}$, allowing the $O_{yz}$ and $O_{zx}$ moments, take place simultaneously by skipping the monoclinic phase. 
For $H \parallel [\bar{1}\, 1\, 0]$, we start from $Cmme$ (No. 67), where the orthorhombic axis is parallel to the field. 
In this case, the hidden order immediately induces the $O_{yz}$ and $O_{zx}$ moments and the tilt of the $c$-axis, resulting in the reduction to $P\bar{1}$ at $T < T_0$. 

With respect to the possibility of quadrupole order, we note a recent article reporting a nonmagnetic phase transition at $T_0=0.4$ K in CeRh$_2$As$_2$ with the CaBe$_2$Ge$_2$-type structure.~\cite{Khim21,Hafner22} 
CeCoSi and CeRh$_2$As$_2$, with the same space group $P4/nmm$, share a similar structural feature, where the site symmetry $4mm$ of Ce ($2c$ site) lacks the inversion symmetry but the crystal is globally centrosymmetric with an inversion center at the midpoint of the two Ce atoms in the unit cell. 
The transition at $T_0=0.4$ K is proposed to be a quadrupole density wave state, involving the conduction electrons through $c$-$f$ hybridization. However, since the $\Gamma_7^{(1)}$ doublet CEF ground state and the $\Gamma_6$ first excited state at $\Delta_{\text{CEF}}\sim 30$ K are well separated, no quadrupolar degree of freedom is expected to remain at 0.4 K. 
A \textit{metaorbital} transition is proposed to be a possible scenario of this ordering, where the CEF excited state is incorporated through a relatively strong hybridization with a Kondo temperature $T_{\text{K}} \sim 30$ K as high as the CEF splitting.~\cite{Hafner22,Hattori10} 
Since the Kondo temperature of CeCoSi, with a localized $4f$ state, is expected to be much lower than that of CeRh$_2$As$_2$, this scenario may not be applicable. 
However, the ratio of $T_0$ to $\Delta_{\text{CEF}}$ of CeCoSi (= 12 K / 120 K) is much larger than that of CeRh$_2$As$_2$ (= 0.4 K / 30 K). 
If we take into account the orbital dependent hybridization, i.e., if the hybridization of the CEF excited state is stronger than that of the ground state, the quadrupole order by incorporating the CEF excited state could be a possible scenario also in CeCoSi. 

Another point worth noting with respect to the structural transition of CeCoSi at $T_0$ above the AFM order is the similar anomaly reported in CeCo$_{1-x}$Fe$_{x}$Si.~\cite{Sereni14,Correa16} 
The increase in the transition temperature by applying a magnetic field is reminiscent of the $T_0$ anomaly in CeCoSi and seems to be pronounced by Fe substitution, although the anomaly is broadened. 
To explain this unusual transition, a phenomenological thermodynamic theory was presented considering the magnetoelastic free energy.~\cite{Oropesa18} 
The theory well explains the field dependences of the structural transitions at $T_{\text{N}}$ and $T_0$, suggesting the importance of magnetoelastic coupling through the $4f$ quadrupole moments. 

A similarity with the structural transition at 13.5 K in CePd$_2$Al$_2$ with the CaBe$_2$Ge$_2$-type structure is also noteworthy.~\cite{Klicpera15} 
The transition temperature increases to $\sim 50$ K by applying a pressure of 3 GPa. The resistivity exhibits a step-like increase reminiscent of a gap opening as in CeCoSi. 
At ambient pressure, the symmetry lowering from tetragonal $P4/nmm$ to orthorhombic $Cmme$ at 13.5 K is accompanied by a significant anomalies in specific heat, magnetic susceptibility, and resistivity, which are much more pronounced than those in CeCoSi possibly because the lattice distortion of $\sim 1$\% is much larger. 
The $4f$ entropy is expected to be involved in this structural transition through a magnetoelastic coupling as demonstrated by the formation of bound states between phonons and CEF excitations.~\cite{Chapon06,Klicpera17} 
The weak anomaly at $T_0$ in CeCoSi may suggest that such effect is weaker than in CePd$_2$Al$_2$.

\section{Summary}
In summary, we have studied the hidden ordered phase of CeCoSi by single-crystal X-ray diffraction in magnetic fields. 
The anomaly at $T_0$ is reflected in the maximum of the $c$-axis lattice parameter. 
At zero field, a structural transition to the triclinic lattice occurs simultaneously with the hidden order at $T_0$. 
For $H \parallel [0\, 1\, 0]$ (equivalent to  $[1\, 0\, 0]$), the temperature of the structural transition, $T_{\text{s1}}$, separates from $T_0$ and decreases with increasing the field, leaving the purely hidden ordered state without triclinic distortion to exist between $T_{\text{s1}}$ and $T_0$. 
A second lattice distortion occurs below $T_{\text{s2}}$, which is slightly lower than $T_{\text{N}}$, leading to a modification of the triclinic lattice reflecting the antiferromagnetic order.  
The preferred structural domains are selected by a weak magnetic field of 1 T. 
The triclinic distortion disappears at high fields above 8 T. 
We also studied the structural transitions for $H \parallel [\bar{1}\, 1\, 0]$ and showed that the phase diagram is much different from that for $H \parallel [0\, 1\, 0]$. 
These results strongly suggest that some electronic order of, e.g, a ferroquadrupole, exists behind and gives rise to a weak magnetic anisotropy in the triclinic phase. However, the origin of the hidden ordered phase, possibly with an antiferro-type order parameter, is still in mystery.

\paragraph{Acknowledgements} 
The authors acknowledge valuable discussions with H. Harima, T. Tayama, H. Hidaka, H. Kotegawa, and M. Manago. 
This work was supported by JSPS KAKENHI Grants No. JP17K05130, 18H03683, JP20H04458, and No. JP20H01854. 
The synchrotron experiments were performed under the approval of the Photon Factory Program Advisory Committee (No. 2020G034). 
The temperature dependences of the lattice parameters from 3 K to 300 K were measured with an X-ray diffractometer (Cu-K$\alpha$) of N-BARD, Hiroshima University.


\clearpage
\renewcommand{\topfraction}{1.0}
\renewcommand{\bottomfraction}{1.0}
\renewcommand{\dbltopfraction}{1.0}
\renewcommand{\textfraction}{0.01}
\renewcommand{\floatpagefraction}{1.0}
\renewcommand{\dblfloatpagefraction}{1.0}
\setcounter{topnumber}{5}
\setcounter{bottomnumber}{5}
\setcounter{totalnumber}{10}
\setcounter{page}{1}

\renewcommand{\theequation}{S\arabic{equation}}
\renewcommand{\thefigure}{S\arabic{figure}}
\renewcommand{\thetable}{S-\Roman{table}}
\setcounter{section}{19}
\setcounter{figure}{0}




\begin{fullfigure}[t]
\begin{center}
\textbf{\large{Supplemental Material}}
\end{center}
\vspace{5mm}

\begin{center}
\textbf{\large{Structural Transition in the Hidden Ordered Phase of CeCoSi}} \\
\vspace{2mm}
T. Matsumura, S. Kishida, M. Tsukagoshi, Y. Kawamura, H. Nakao, and H. Tanida
\end{center}
\vspace{20mm}

\begin{center}
\includegraphics[width=5cm]{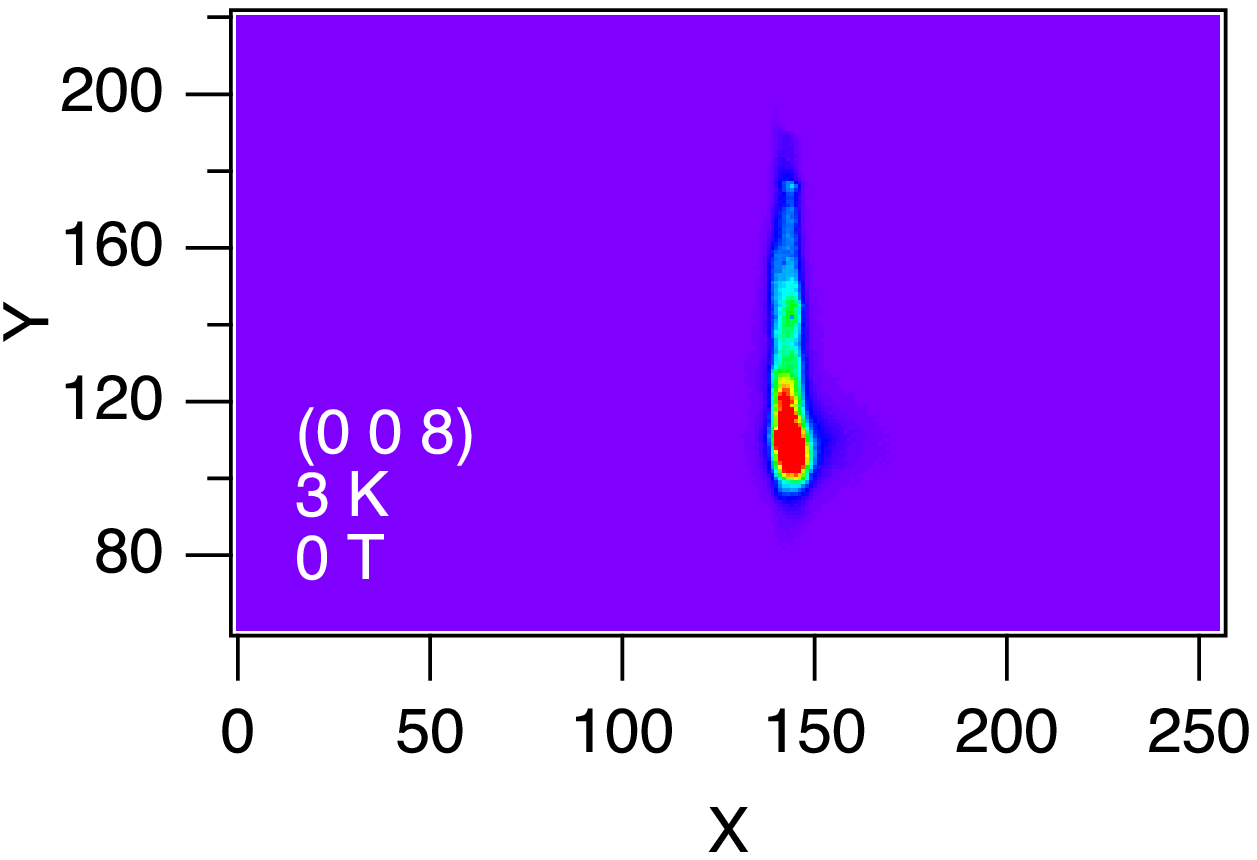}
\includegraphics[width=5cm]{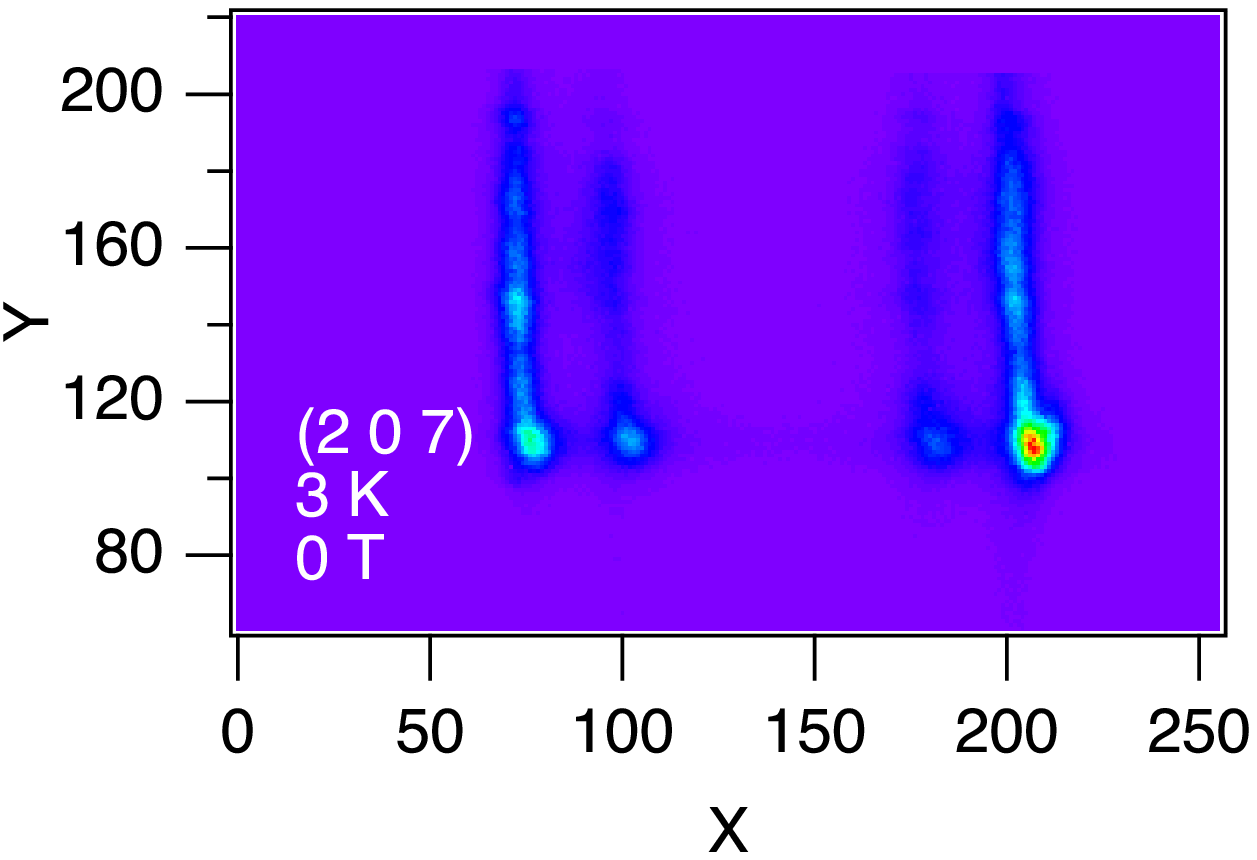}
\includegraphics[width=5cm]{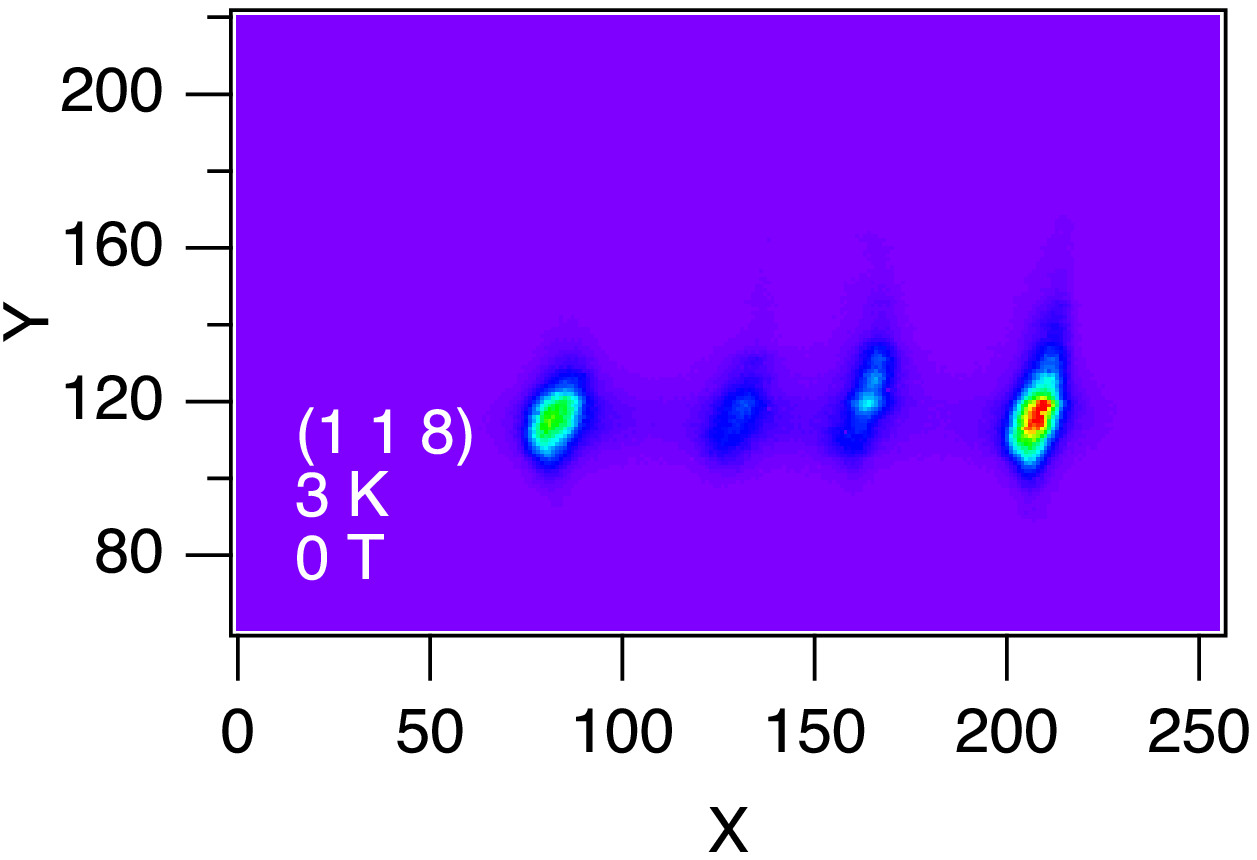}\\
\includegraphics[width=5cm]{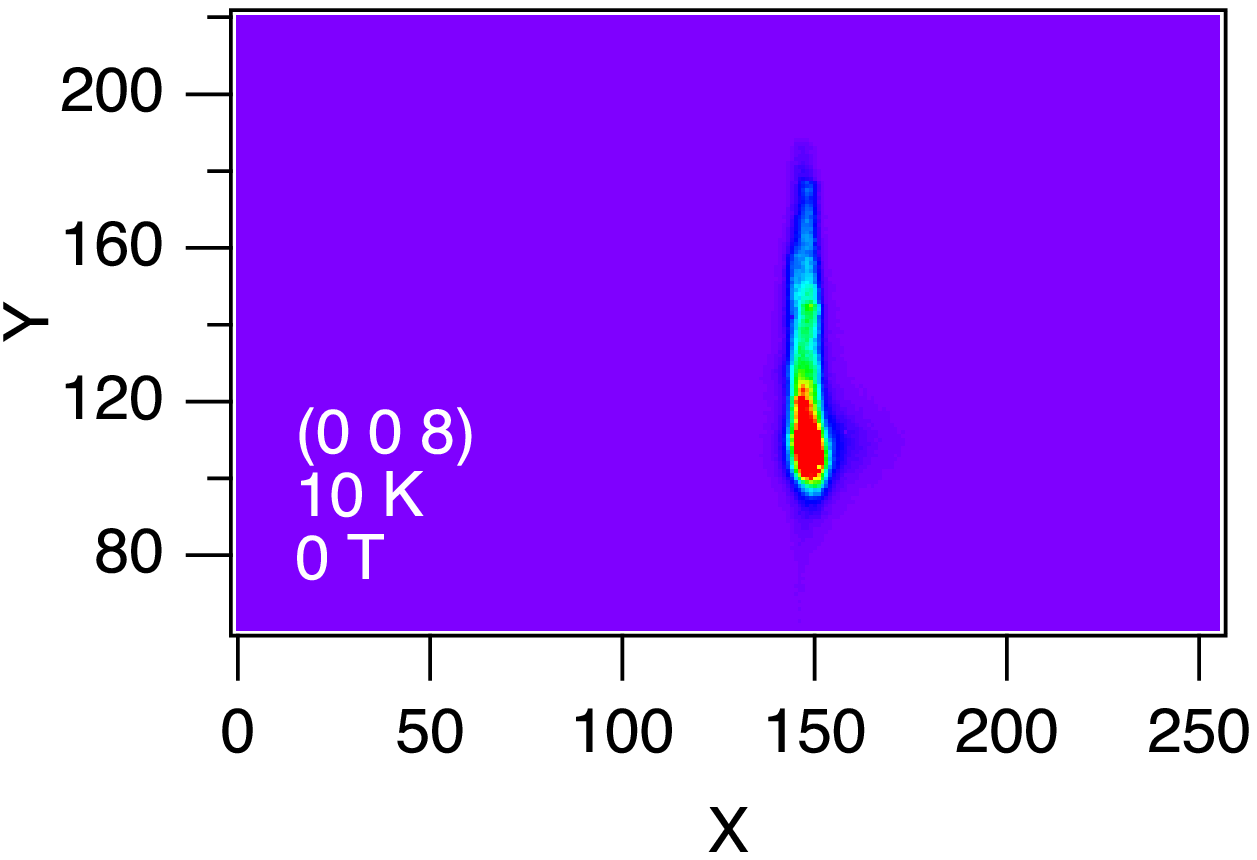}
\includegraphics[width=5cm]{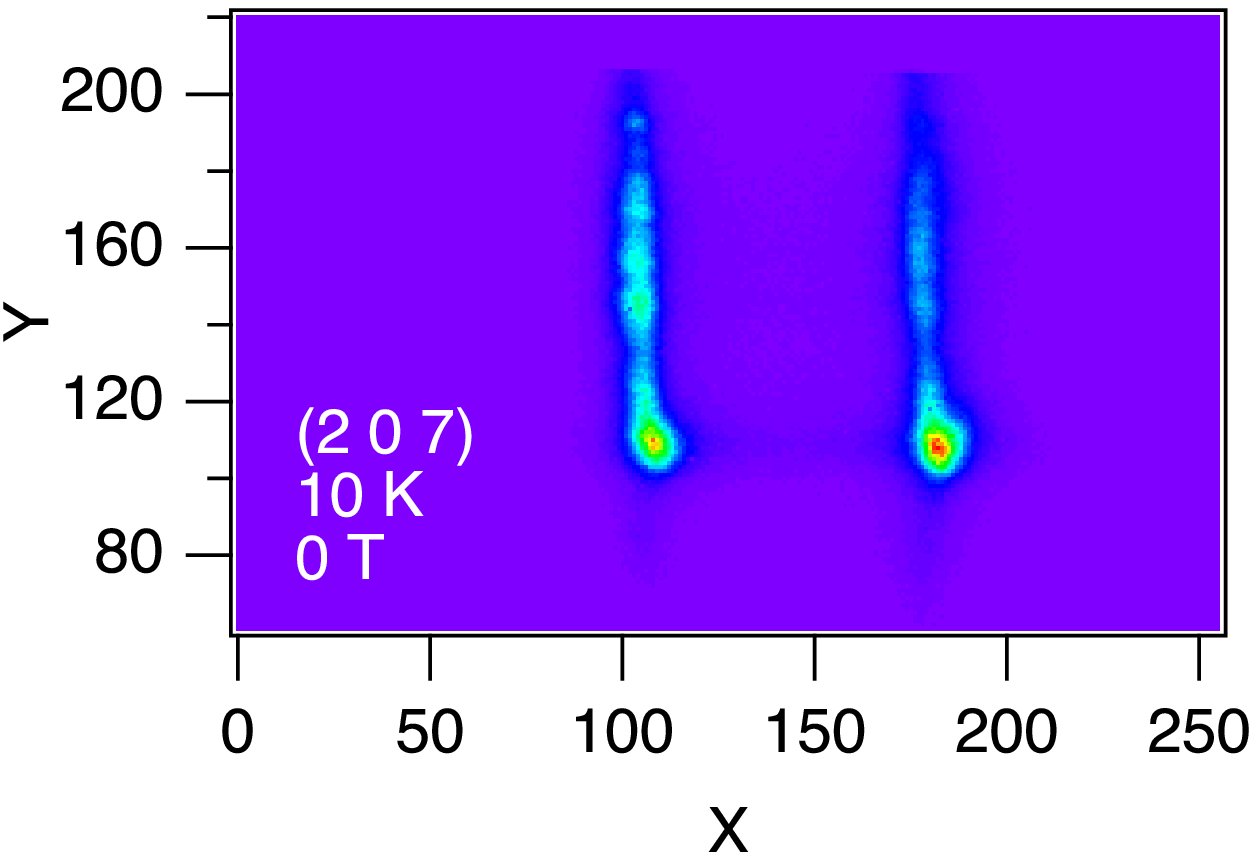}
\includegraphics[width=5cm]{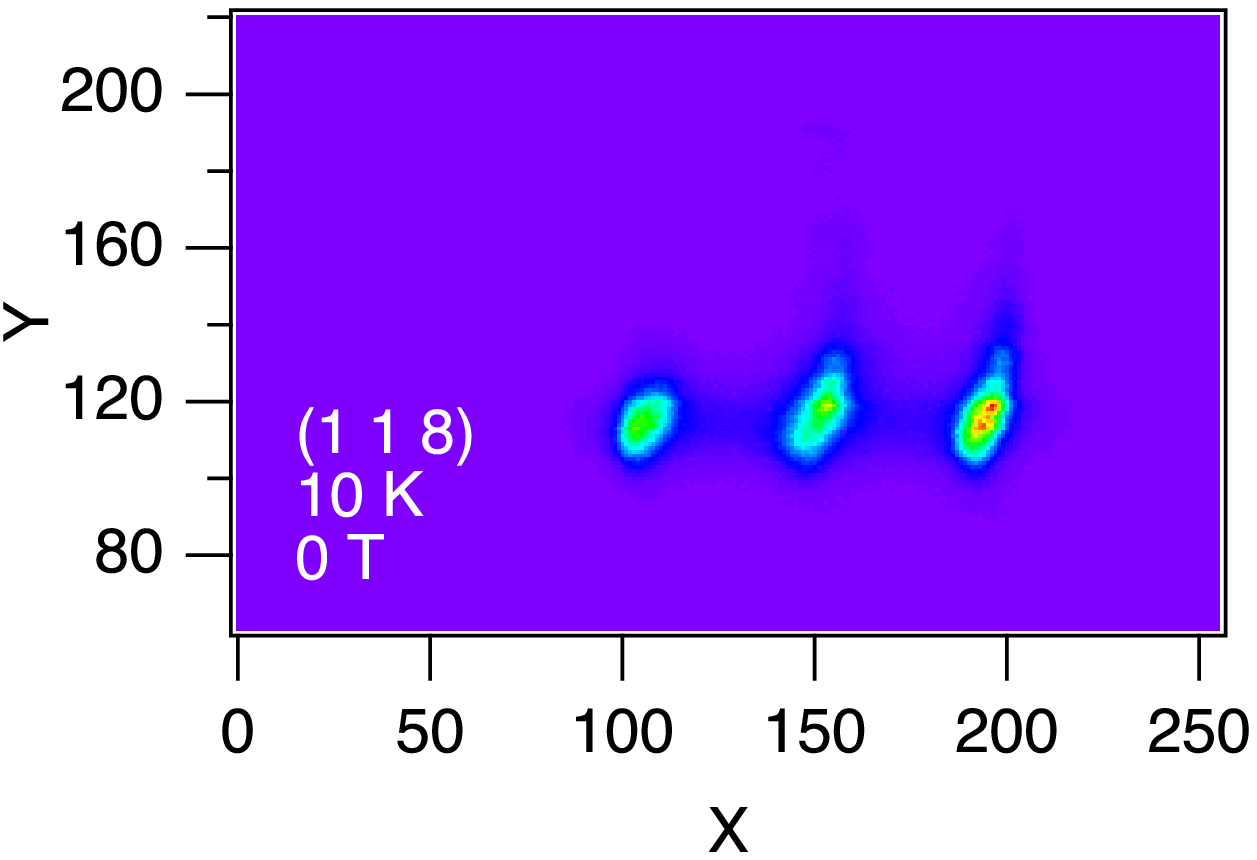}
\caption{
X-ray oscillation photographs of the (0 0 8), (2 0 7), and (1 1 8) reflections at 3 K and 10 K at zero field. 
X and Y represent the pixel number of the detector. 
The intensity sum along Y between 90 and 130 were used to obtain the one dimensional $2\theta$ dependence of the intensity (Fig. \ref{fig:tthscans0T} in the main text. Figs. \ref{fig:tthscans008}, \ref{fig:tthscans207}, \ref{fig:tthscans118}, \ref{fig:tthscans3K207}, \ref{fig:tthscans3K118}, and  \ref{fig:tthscans10K118} in the supplemental material). 
One pixel along X corresponds to the $2\theta$ angle of $0.0069^{\circ}$. 
The crystal angle ($\theta$) was rotated by approximately two degrees about the Bragg peak. 
The $\theta$-scan was also performed to determine the peak position and deduce the $(H, K, L)$ index of the split peaks (Figs. \ref{fig:Tdep008L}, \ref{fig:Tdep207L}, \ref{fig:Hdep207L}, \ref{fig:Tdep118L}, and \ref{fig:Hdep118L} in the main text. Figs. \ref{fig:Tdep207H} and \ref{fig:Tdep118H} in the supplemental material).  
}
\label{fig:imageOSC}
\end{center}
\end{fullfigure}

\begin{fullfigure}[h]
\begin{center}
\includegraphics[width=15cm]{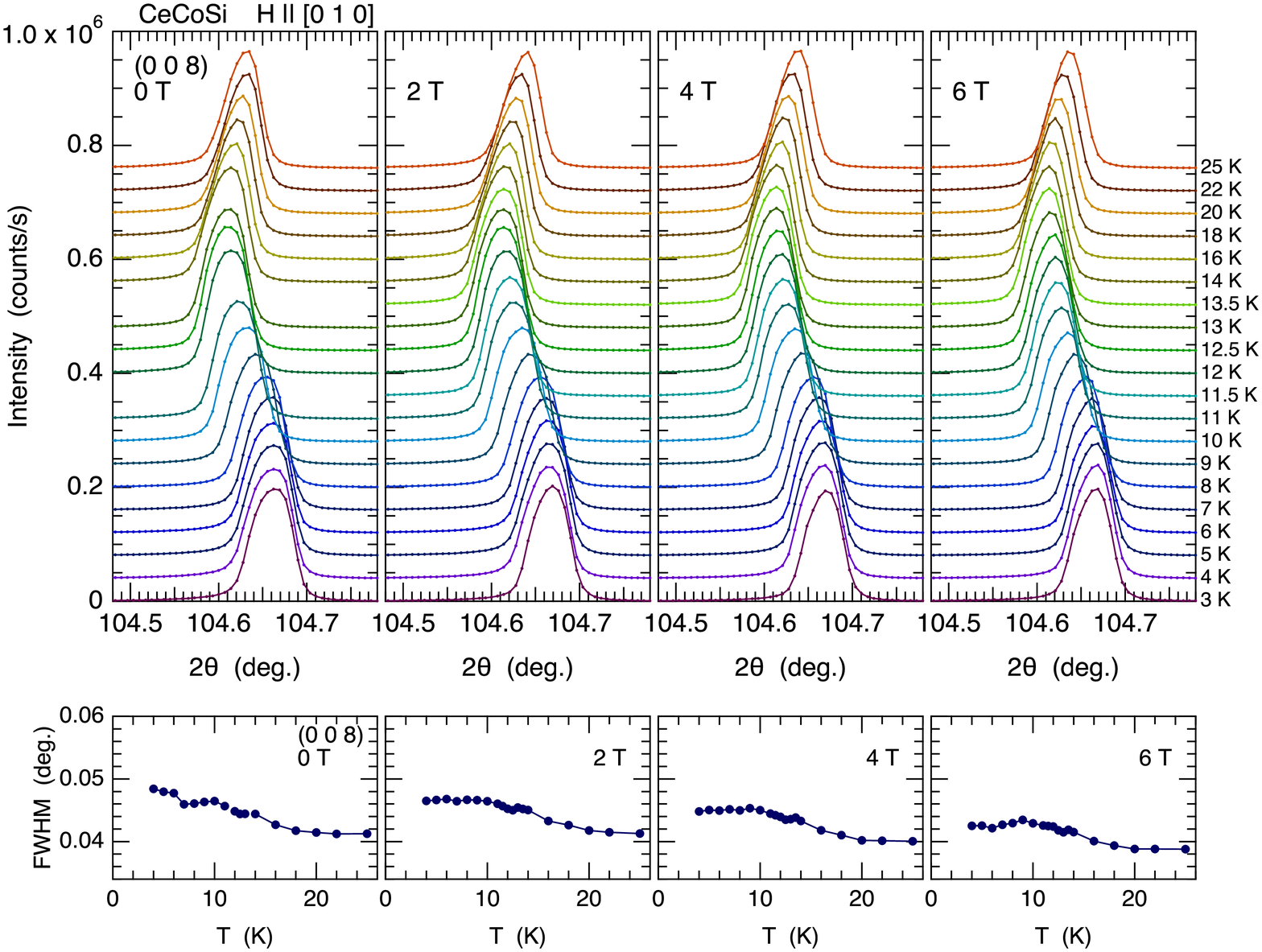}
\caption{
(top) Scattering angle ($2\theta$) dependences of the intensity of the (0 0 8) reflection in magnetic fields of 0 T, 2 T, 4 T, and 6 T applied along $[0\, 1\, 0]$. 
The $T$-dependences of $L$ in $(H, K, L)$ for (0 0 8) shown in Fig.~\ref{fig:Tdep008L} in the main text are obtained from the peak positions of these data. 
(bottom) $T$-dependences of the full width at half maximum (FWHM) of the Bragg peak obtained from the fit with a squared Lorentzian function. 
For (0 0 8), no cusp anomaly is observed in the FWHM, except for the gradual variation in slope at around 15 K. 
}
\label{fig:tthscans008}
\end{center}
\end{fullfigure}

\begin{fullfigure}[h]
\begin{center}
\includegraphics[width=15cm]{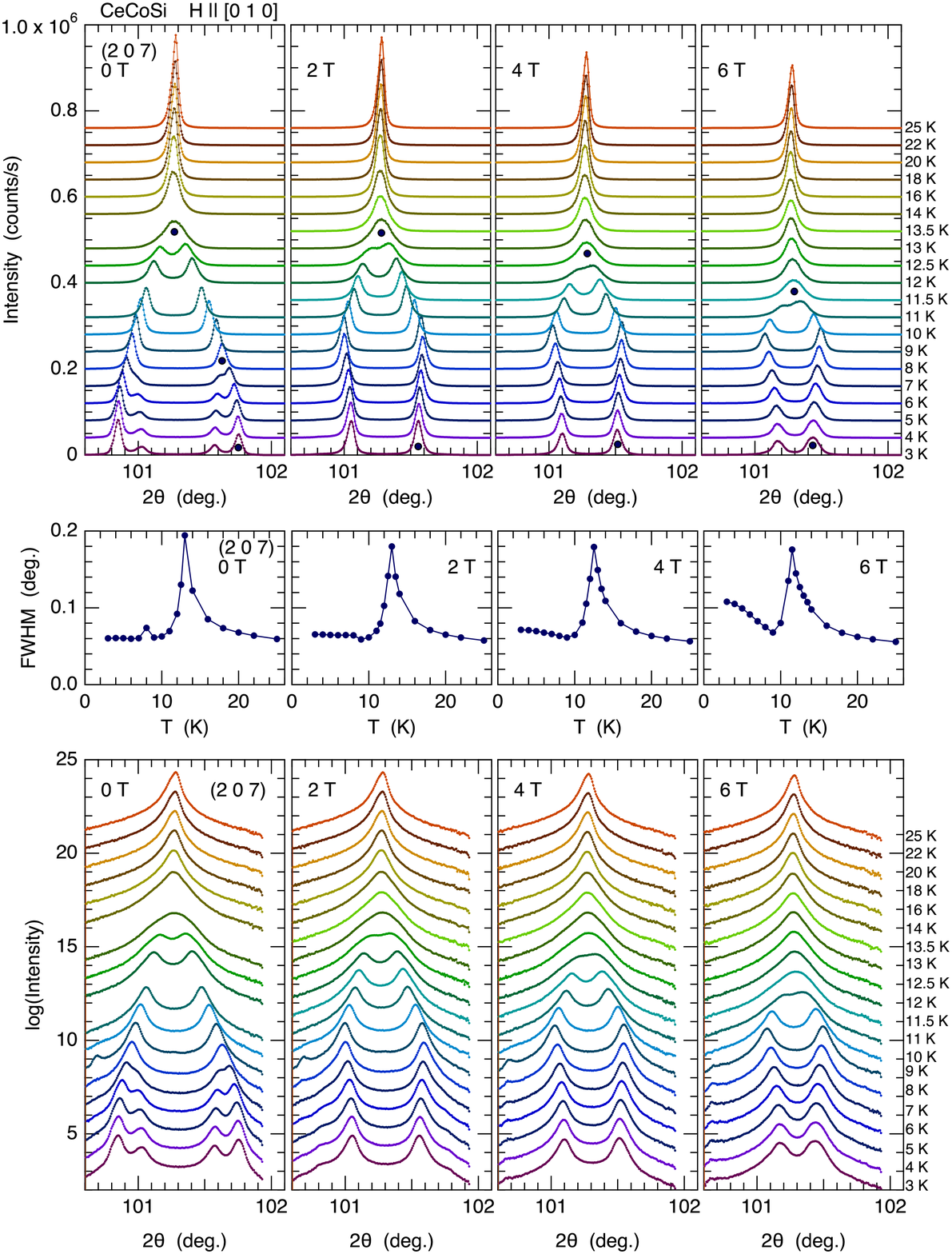}
\caption{
(top) Scattering angle ($2\theta$) dependences of the intensity of the (2 0 7) reflection in magnetic fields of 0 T, 2 T, 4 T, and 6 T applied along $[0\, 1\, 0]$. 
The $T$-dependences of $L$ in $(H, K, L)$ for (2 0 7) shown in Fig.~\ref{fig:Tdep207L} in the main text are obtained from the peak positions of these data. 
(middle) $T$-dependences of the full width at half maximum (FWHM) of the Bragg peak obtained from the fit with multiple squared Lorentzian functions. 
The FWHM is shown for the peak indicated by the mark (closed circle) in the top panels. 
The peak profiles for $T\ge 13$ K, 13 K, 12.5 K, and 11.5 K for 0 T, 2 T, 4 T, and 6 T, respectively, which are indicated by the marks in the top panels, were treated as a single peak. Below these temperatures the profiles were treated as double peaks. 
For 0 T, the profiles for $8  \le T \le 12.5$ K and for $T \le 7$ K were treated as double peaks and four peaks, respectively, which are also indicated by the marks. 
(bottom) Scattering angle ($2\theta$) dependences of the intensity plotted in a logarithmic scale.  
}
\label{fig:tthscans207}
\end{center}
\end{fullfigure}

\begin{fullfigure}[h]
\begin{center}
\includegraphics[width=15cm]{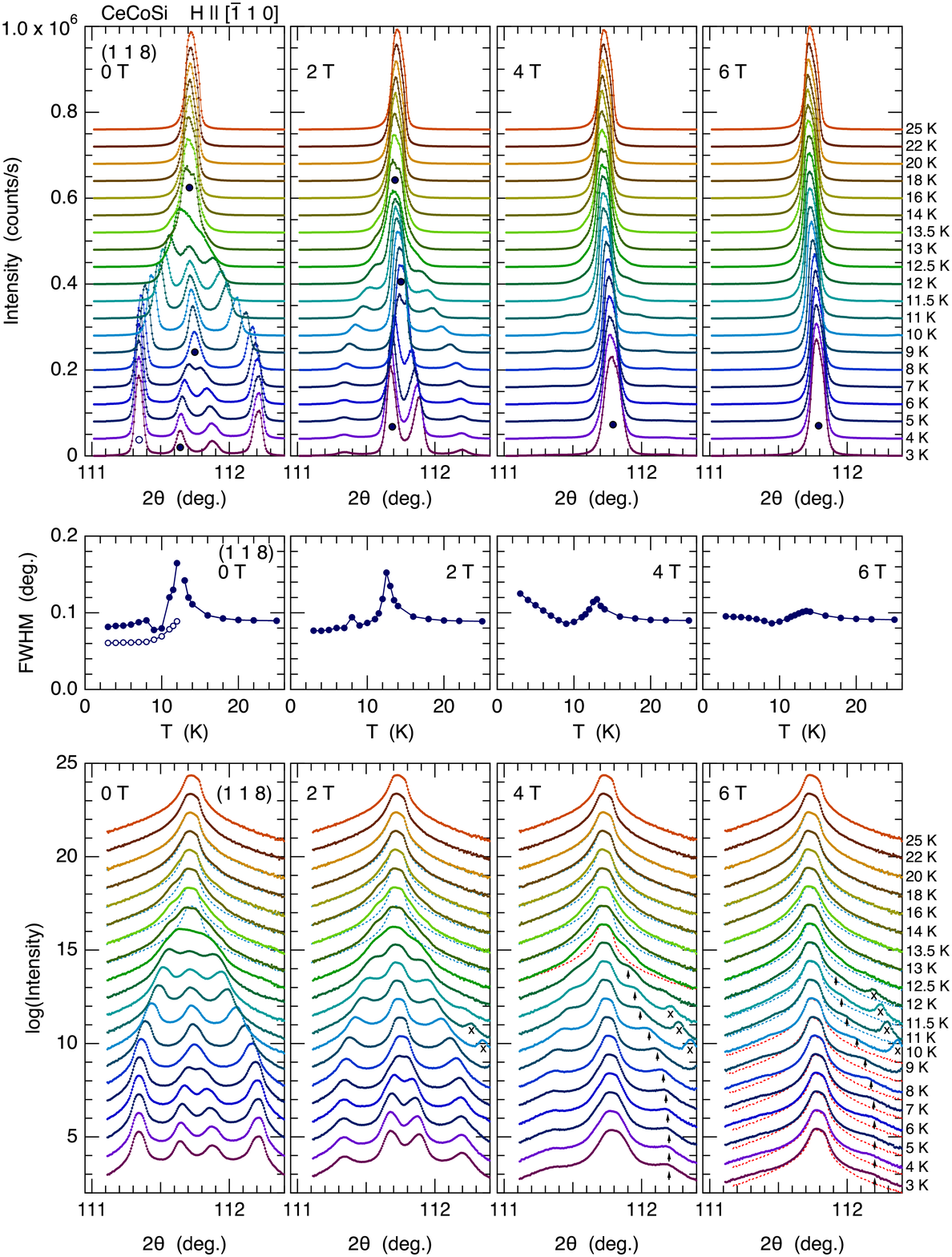}
\caption{
(top) Scattering angle ($2\theta$) dependences of the intensity of the (1 1 8) reflection in magnetic fields of 0 T, 2 T, 4 T, and 6 T applied along $[\bar{1}\, 1\, 0]$. 
(middle) $T$-dependences of the full width at half maximum (FWHM) of the Bragg peak obtained from the fit with multiple squared Lorentzian functions. 
The FWHM is shown for the peak indicated by the mark (closed circle) in the top panels. For 0 T, the FWHM for another peak (open circle) is also shown, which does not split below $T_{\text{s2}}$. 
The peak profiles for $T\ge 13$ K for 0 T and  2 T, and all the profiles for 4 T and 6 T, were treated as a single peak as indicated by the marks in the top panels. 
For 0 T and 2 T, the profiles for $8  \le T \le 12.5$ K and for $T \le 7$ K were treated as three peaks (one central and two outer peaks) and four peaks (two central and two outer peaks) , respectively. 
The broad profile at 12.5 K for 0 T, which consists of three unresolved peaks, was excluded from the FWHM analysis. 
(bottom) Scattering angle ($2\theta$) dependences of the intensity plotted in a logarithmic scale.  The dashed lines are the profile at 25 K for comparison. The arrows in the panels of 4 T and 6 T indicate the positions of the weakly observed outer peaks. The x marks indicate  the unidentified peaks. 
}
\label{fig:tthscans118}
\end{center}
\end{fullfigure}

\begin{fullfigure}[t]
\begin{center}
\includegraphics[width=7.5cm]{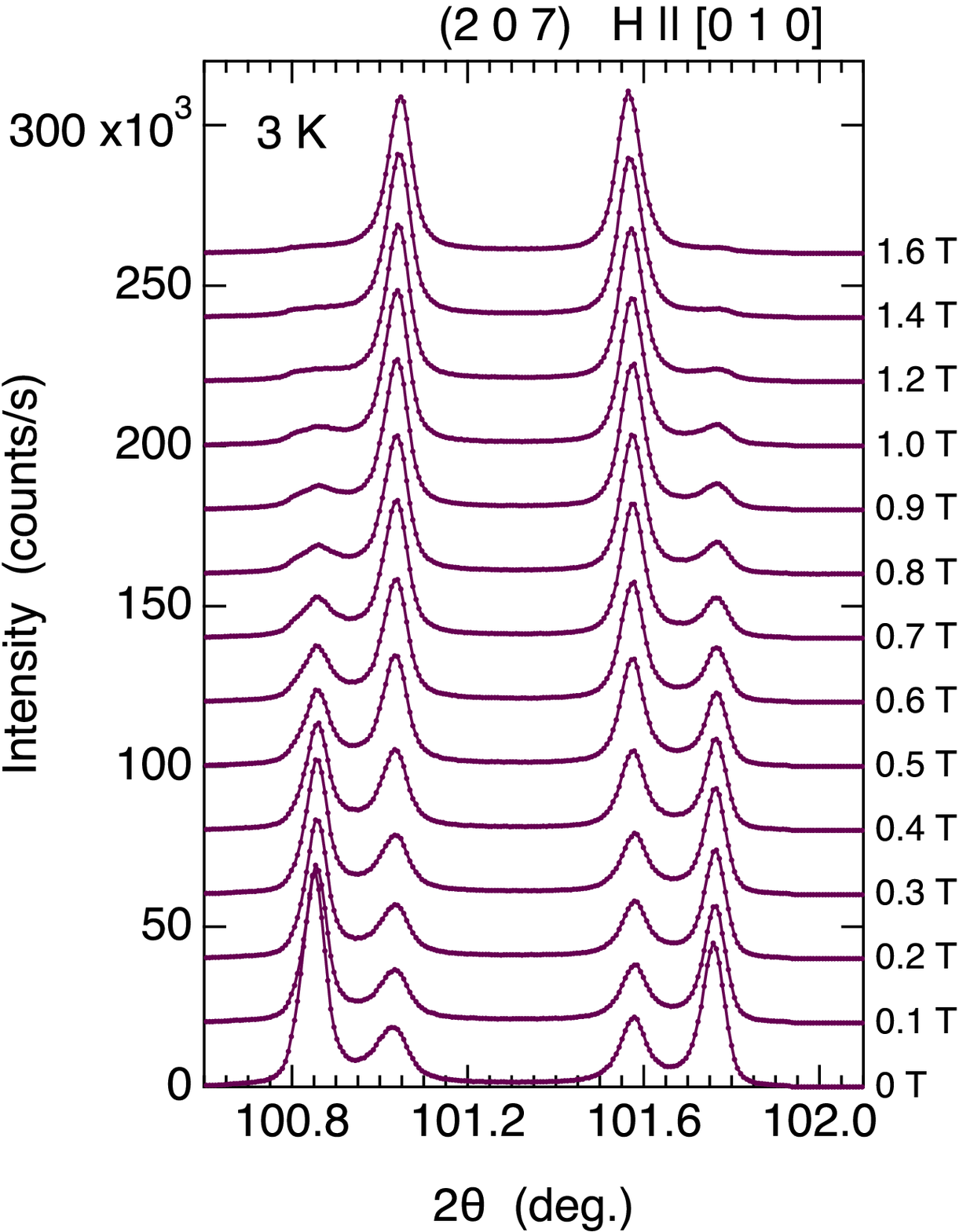}
\caption{
Scattering angle ($2\theta$) dependences of the intensity of the (2 0 7) reflection at 3 K in magnetic fields applied along $[0\, 1\, 0]$. 
The intensities of the outer peaks rapidly decrease by applying the field and almost disappear above 2 T.  
The intensities transfer from the outer peaks to the inner peaks. 
The field dependence of $L$ in $(H, K, L)$ for (2 0 7) shown in Fig.~\ref{fig:Hdep207L} in the main text are obtained from these data. 
}
\label{fig:tthscans3K207}
\end{center}
\end{fullfigure}

\begin{fullfigure}[t]
   \begin{minipage}[b]{0.47\textwidth}
       \begin{center}
       \includegraphics*[width=7.5cm]{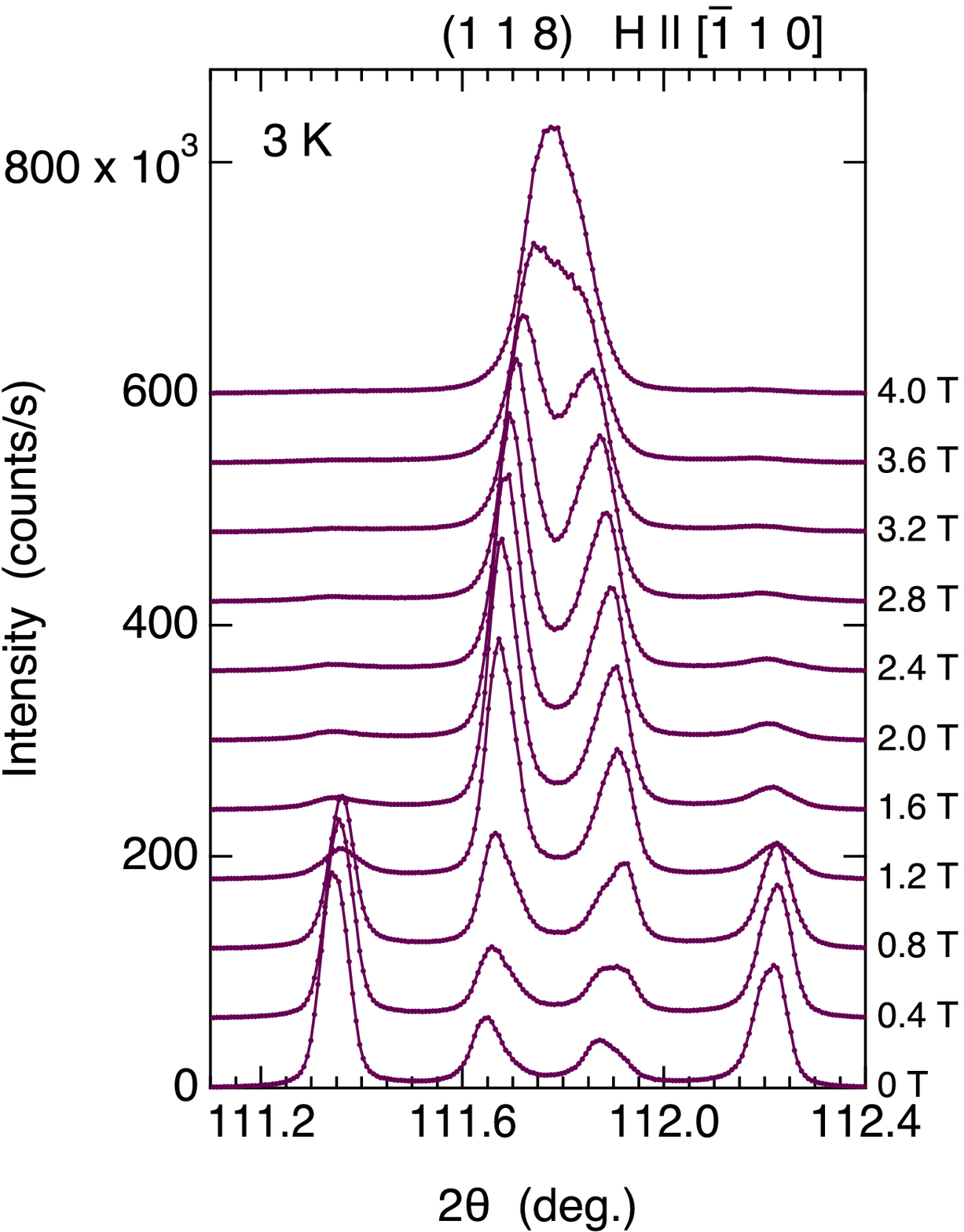}
       \end{center}
\caption{
Scattering angle ($2\theta$) dependences of the intensity of the (1 1 8) reflection at 3 K in magnetic fields applied along $[\bar{1}\, 1\, 0]$. 
The intensities of the outer peaks abruptly decrease at 1 T and transfer to the inner peaks. 
The width of the peak split between the inner peaks gradually decrease and merge back into the central peak above 4 T. 
The intensities of the outer peaks weakly remain up to 4 T, which seems to remain even at 6 T very weakly (see the logarithmic plot in Fig.~\ref{fig:tthscans118}). 
}
\label{fig:tthscans3K118}
   \end{minipage}
\hfill
   \begin{minipage}[b]{0.47\textwidth}
       \begin{center}
       \includegraphics*[width=8cm]{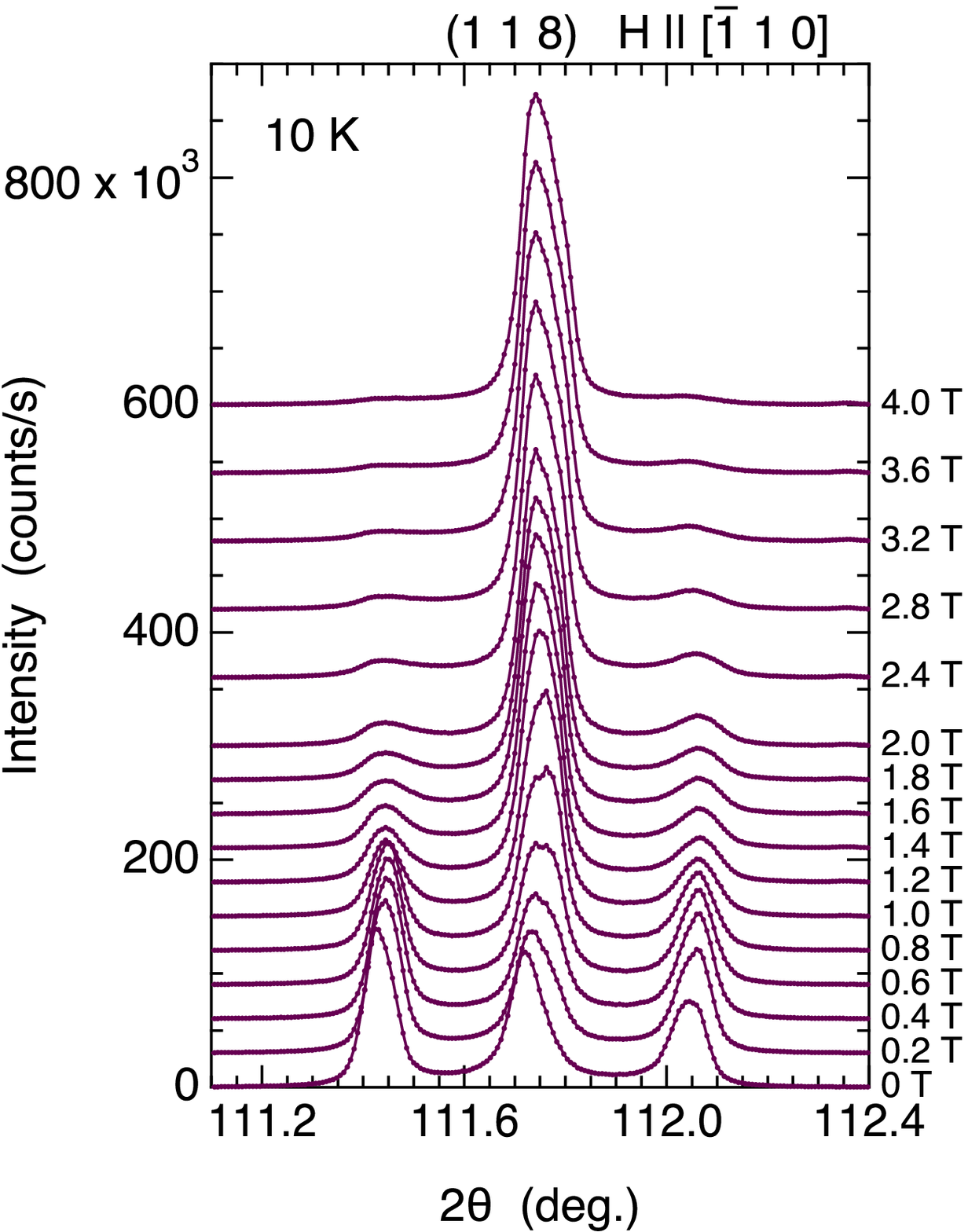}
       \end{center}
\caption{
Scattering angle ($2\theta$) dependences of the intensity of the (1 1 8) reflection at 10 K in magnetic fields applied along $[\bar{1}\, 1\, 0]$. 
The intensities of the outer peaks exhibit a steep decrease at around 1 T and a gradual decrease up to 4 T. 
The decreased intensity is transferred to the central peak. 
The intensities of the outer peaks weakly remain up to 4 T, which seems to remain even at 6 T very weakly (see the logarithmic plot in Fig.~\ref{fig:tthscans118}). 
The field dependence of $L$ in $(H, K, L)$ for (1 1 8) shown in Fig.~\ref{fig:Hdep118L} in the main text are obtained from these data of Figs.~\ref{fig:tthscans3K118} and \ref{fig:tthscans10K118}. 
}
\label{fig:tthscans10K118}
   \end{minipage}
\end{fullfigure}

\begin{fullfigure}[t]
   \begin{minipage}[b]{0.47\textwidth}
       \begin{center}
       \includegraphics[width=8cm]{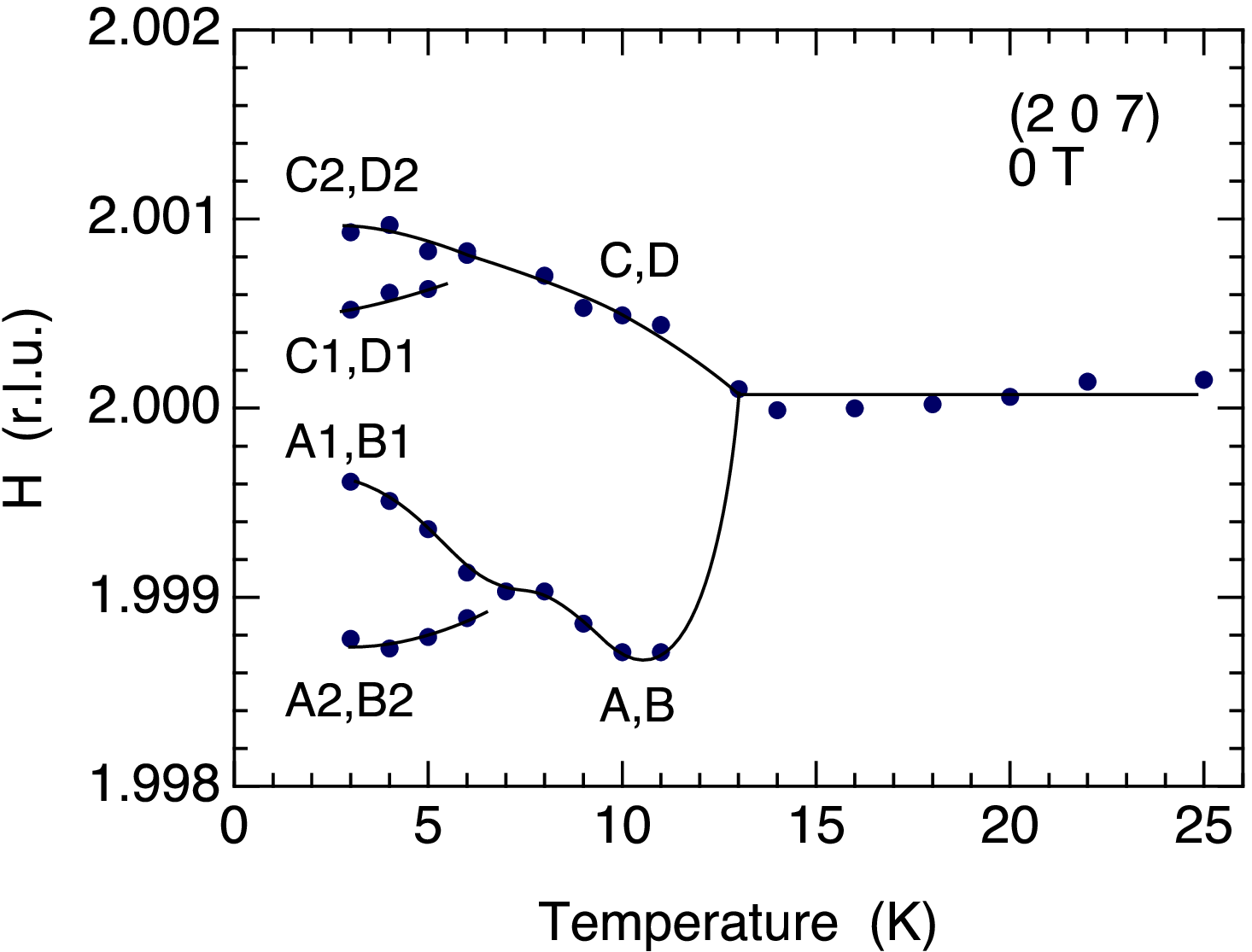}
       \end{center}
\caption{
Temperature dependence of the peak position of the (2 0 7) reflection, $H$ in $(H, K, L)$, expressed in the reciprocal lattice unit of the tetragonal lattice. 
The labels represent the domain assignments shown in Fig.~\ref{fig:triclinic} in the main text. 
The asymmetric and complex behavior of the peak position may be ascribed to the subtle change in the sample orientation due to the structural deformation. We took into account only the magnitude of the peak split to estimate the parameters of triclinic distortion. 
}
\label{fig:Tdep207H}
   \end{minipage}
\hfill
   \begin{minipage}[b]{0.47\textwidth}
       \begin{center}
       \includegraphics*[width=8cm]{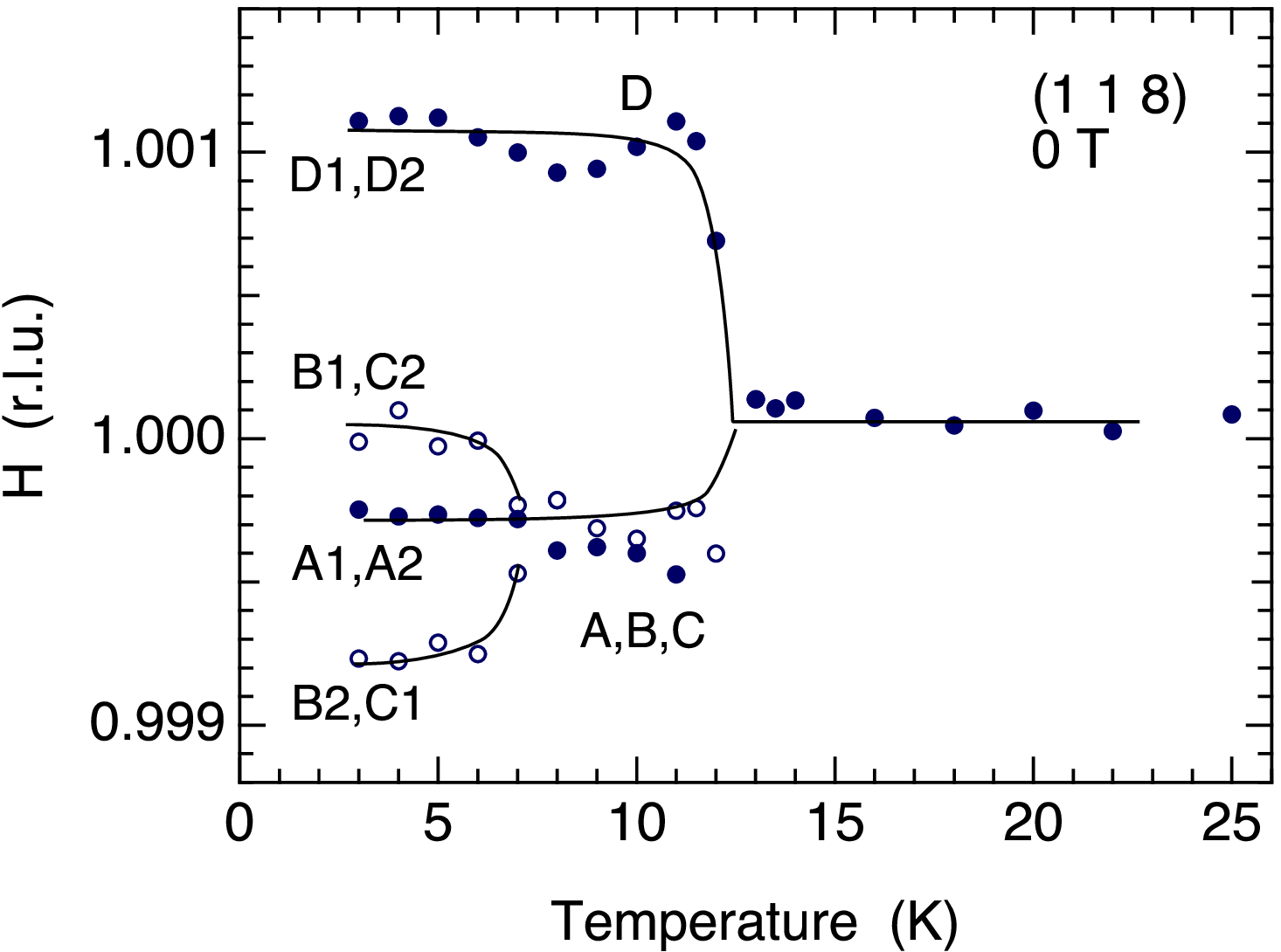}
       \end{center}
\caption{
Temperature dependence of the peak position of the (1 1 8) reflection, $H$ in $(H, K, L)$, expressed in the reciprocal lattice unit of the tetragonal lattice. 
The labels represent the domain assignments shown in Fig.~\ref{fig:triclinic} in the main text. 
The asymmetric behavior of the peak position may be ascribed to the subtle change in the sample orientation due to the structural deformation. 
}
\label{fig:Tdep118H}
   \end{minipage}
\end{fullfigure}

\begin{fullfigure}[t]
\begin{center}
\includegraphics[width=8cm]{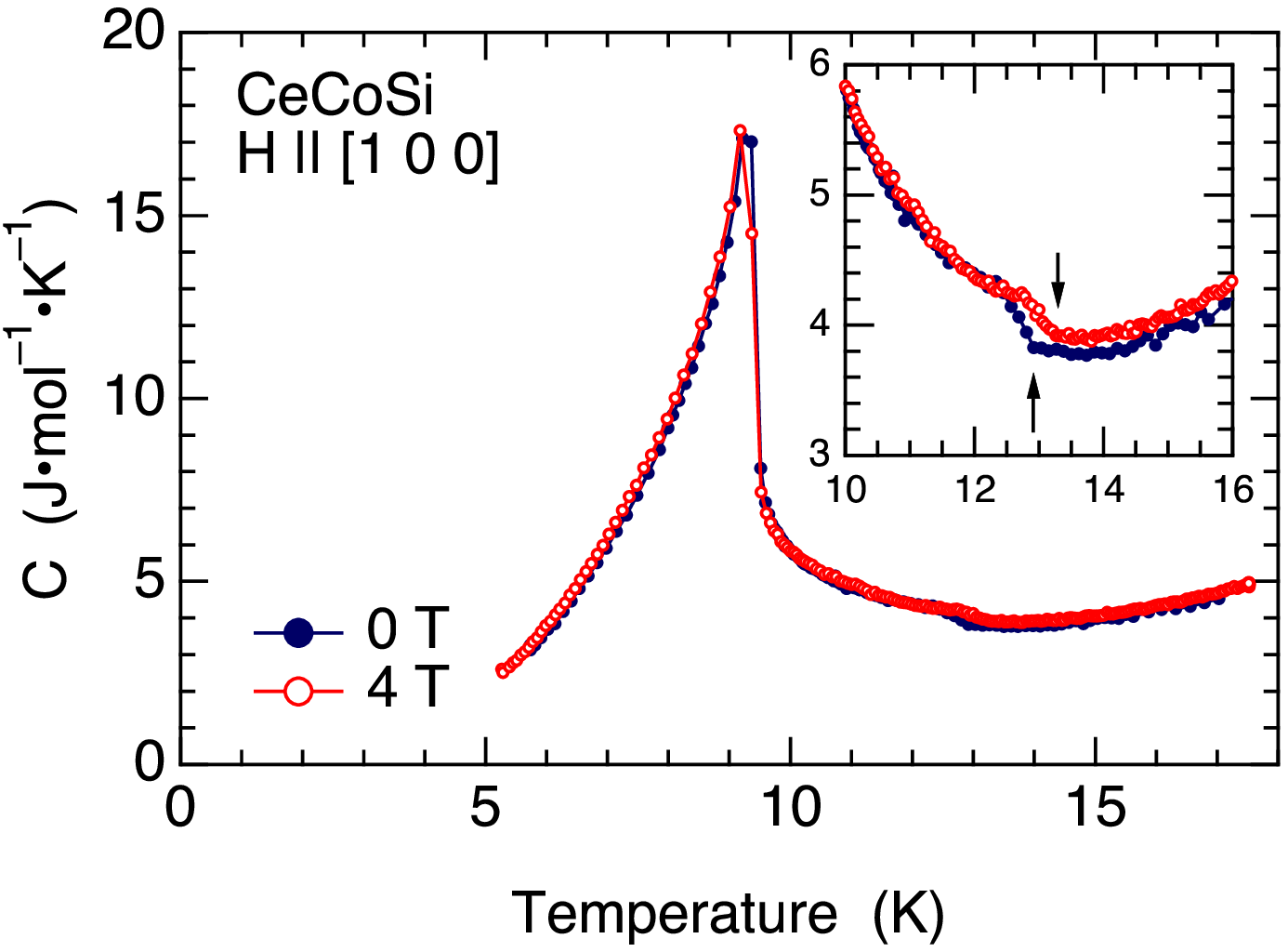}
\caption{
Temperature dependence of the specific heat of the sample used in the X-ray diffraction in this study. The arrows in the inset indicate the anomaly at $T_0$. 
}
\label{fig:Spch}
\end{center}
\end{fullfigure}

\begin{fullfigure}[t]
\begin{center}
\includegraphics[width=15cm]{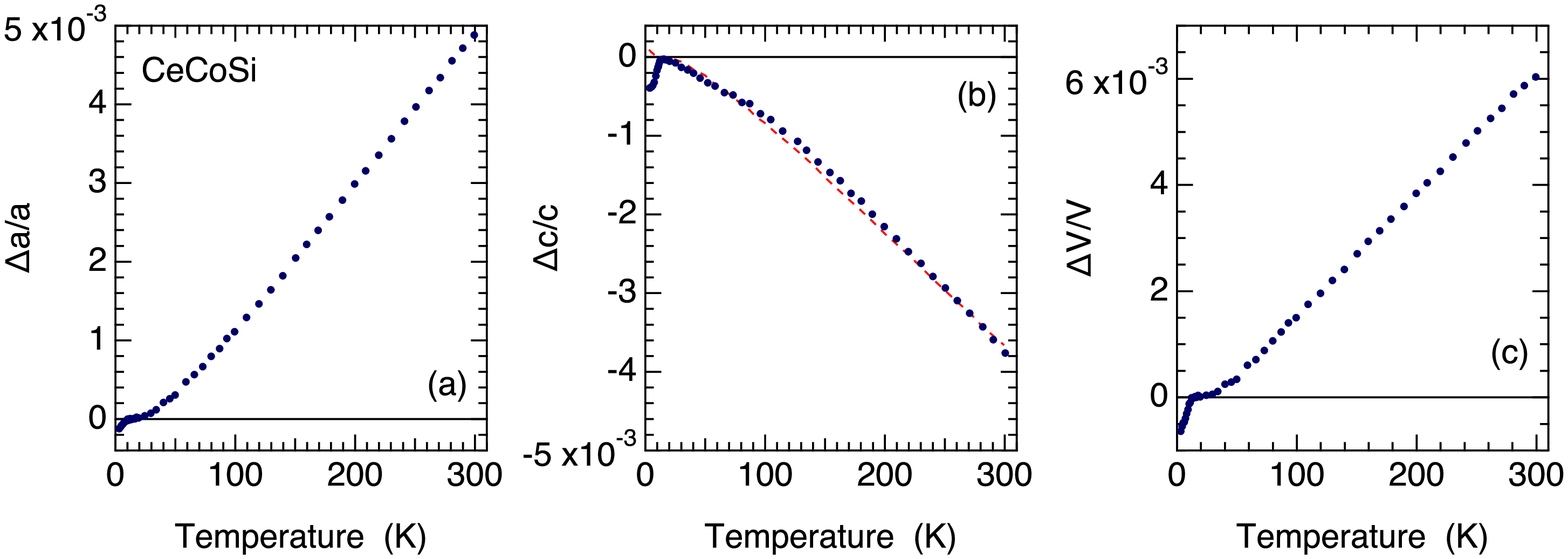}
\caption{
Temperature dependences of the relative change in the lattice parameters (a) $\Delta a/a$ and (b) $\Delta c/c$. 
$a=4.057$ \AA\ and $c=6.987$ \AA\ at room temperature [H. Tanida \textit{et al.}, J. Phys. Soc. Jpn. \textbf{88}, 054716 (2019)]. 
The lattice parameters were measured by using a laboratory based X-ray diffractometer with a Cu target.  
The $\Delta a/a$ and $\Delta c/c$ data were deduced from (4 0 0) and (0 0 8) reflections, respectively. 
The dashed line in (b) represents $-0.75(\Delta a/a)$. 
No splitting was observed in the (4 0 0) reflection because of the large crystal mosaicity of the side edge of the sample used in the measurement: 
plate-shaped with an as-grown $c$-plane surface, 1.5 $\times$ 2.0 mm$^2$ in area and 0.3 mm in thickness. 
(c) Temperature dependence of the relative change in volume of the tetragonal unit cell. 
}
\label{fig:dadcdV}
\end{center}
\end{fullfigure}

\begin{fullfigure}[t]
\begin{center}
\includegraphics[width=18cm]{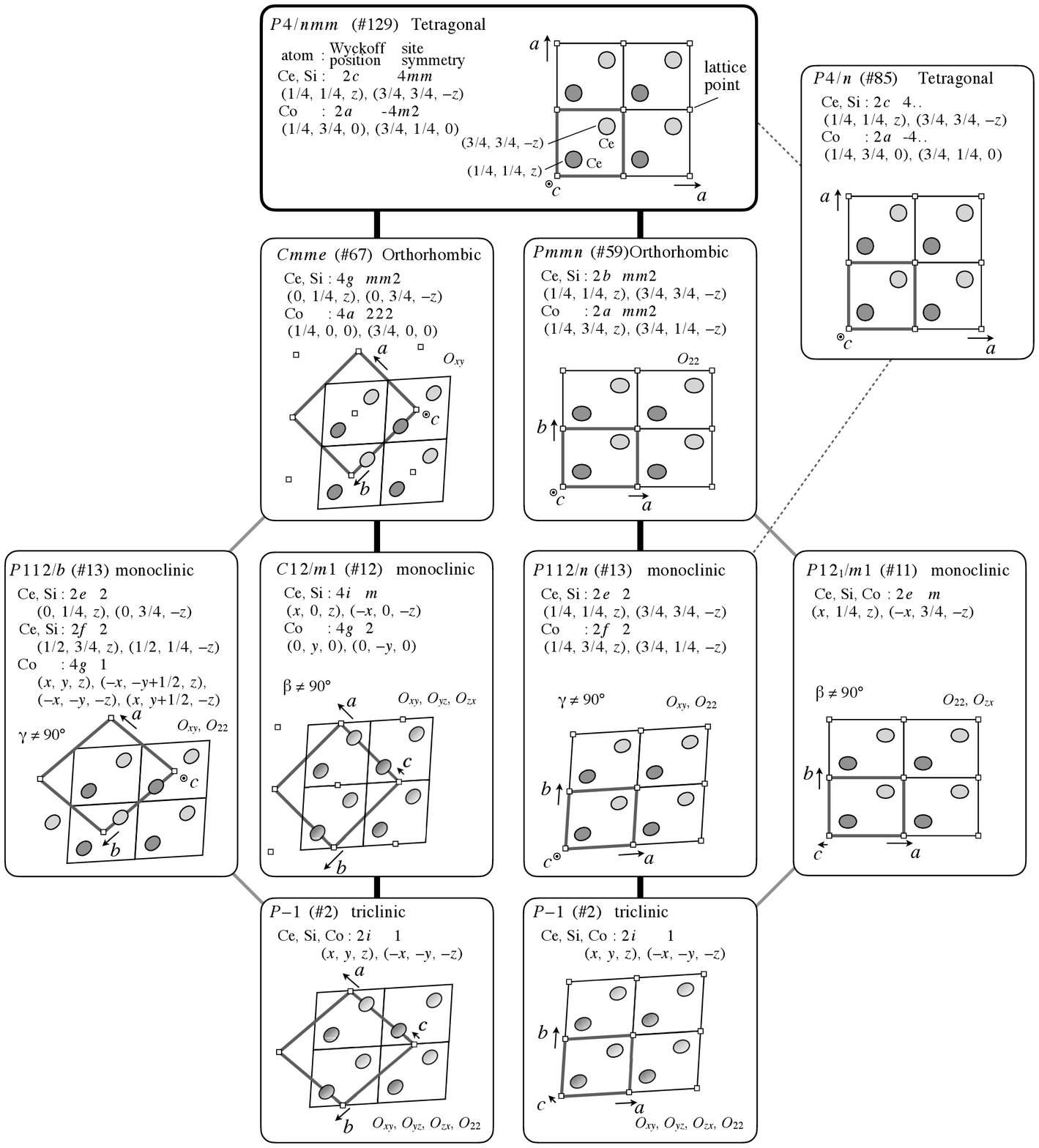}
\vspace{8mm}
\caption{
Group-subgroup relations of CeCoSi from $P4/nmm$ (\# 129) to $P\bar{1}$ (\# 2) [International Tables for Crystallography, ed. T. Hahn (Springer,
Heidelberg, 2005) Vol. A]. 
The thick black lines represent the likely processes we discussed in the main text. 
Ellipsoids schematically represent the quadrupole moments of Ce. 
Thick gray parallelograms represent the unit cell. Note that $Cmme$ (\# 67) and $C12/m1$ (\# 12) are C-centered. For $H \parallel [0\,1\,0]$, we proposed a possibility of the symmetry reduction from $Pmmn$ (\# 59) at $T>T_0$, to $P112/n$ (\# 13) at $T_{\text{s1}} < T < T_0$, and to $P\bar{1}$ (\# 2) at $T < T_{\text{s1}}$. For $H \parallel [\bar{1}\, 1\, 0]$, we proposed that it starts from $Cmme$ (\# 67) at $T > T_0$ and is immediately reduced to $P\bar{1}$ (\# 2) at $T < T_0$. The gray lines are the ones we consider unlikely. $P4/n$ (\# 85), connected by the gray dashed lines, is not likely because the mirror symmetry is lost, which is not the case here. $P 1 2_1/m 1$ (\# 11) for $H \parallel [0\, 1\, 0]$ and $P 1 1 2/b$ (\# 13) for $H \parallel [\bar{1}\, 1\, 0]$ are unlikely because the experiment shows that the $c$-axis tilts toward the $[1\, 1\, 0]$ direction. }
\label{fig:subgroupCeCoSi}
\end{center}
\end{fullfigure}

\end{document}